\documentclass[sigconf]{acmart}

\usepackage{multirow}
\usepackage{multicol}
\usepackage{enumitem}
\usepackage{graphicx}
\usepackage{subcaption}
\usepackage{tabularx}
\usepackage{bbm}
\usepackage{stmaryrd}

\AtBeginDocument{%
  \providecommand\BibTeX{{%
    \normalfont B\kern-0.5em{\scshape i\kern-0.25em b}\kern-0.8em\TeX}}}

\copyrightyear{2024}
\acmYear{2024}
\setcopyright{acmlicensed}\acmConference[KDD '24]{Proceedings of the 30th
ACM SIGKDD Conference on Knowledge Discovery and Data Mining}{August
25--29, 2024}{Barcelona, Spain}
\acmBooktitle{Proceedings of the 30th ACM SIGKDD Conference on Knowledge
Discovery and Data Mining (KDD '24), August 25--29, 2024, Barcelona, Spain}
\acmDOI{10.1145/3637528.3671523}
\acmISBN{979-8-4007-0490-1/24/08}

\begin{document}

\title{Residual Multi-Task Learner for Applied Ranking}

\author{Cong Fu}
\affiliation{
 \institution{Shopee Pte. Ltd.}
 \city{Singapore}
 \country{Singapore}
}
\email{fc731097343@gmail.com}

\author{Kun Wang}
\affiliation{
 \institution{Shopee Pte. Ltd.}
 \city{Shanghai}
 \country{China}
}
\email{wk1135256721@gmail.com}

\author{Jiahua Wu}
\affiliation{
 \institution{Shopee Pte. Ltd.}
 \city{Singapore}
 \country{Singapore}
}
\email{gauvain.wujiahua@gmail.com}

\author{Yizhou Chen}
\affiliation{
 \institution{Shopee Pte. Ltd.}
 \city{Singapore}
 \country{Singapore}
}
\email{yizhou.chen@shopee.com}

\author{Guangda Huzhang}
\affiliation{
 \institution{Shopee Pte. Ltd.}
 \city{Singapore}
 \country{Singapore}
}
\email{guangda.huzhang@shopee.com}

\author{Yabo Ni}
\affiliation{
 \institution{Nanyang Technological University} 
 \city{Singapore}
 \country{Singapore}
}
\email{yabo001@e.ntu.edu.sg}

\author{Anxiang Zeng}
\affiliation{
 \institution{SCSE, Nanyang Technological University}
 \city{Singapore}
 \country{Singapore}
}
\email{zeng0118@e.ntu.edu.sg}

\author{Zhiming Zhou}
\authornote{Corresponding author.}
\affiliation{
 \institution{ECONCS\footnote{}\authornote{Key Laboratory of Interdisciplinary Research of Computation and Economics.}, Shanghai University of Finance and Economics}
 % \institution{Key Laboratory of Interdisciplinary Research of Computation and Economics, Shanghai University of Finance and Economics}
 \city{Shanghai}
 \country{China}
}
\email{zhouzhiming@mail.shufe.edu.cn}

\renewcommand{\shortauthors}{Cong Fu et al.}

\begin{abstract}
    Modern e-commerce platforms rely heavily on modeling diverse user feedback to provide personalized services. Consequently, multi-task learning has become an integral part of their ranking systems. However, existing multi-task learning methods encounter two main challenges: some lack explicit modeling of task relationships, resulting in inferior performance, while others have limited applicability due to being computationally intensive, having scalability issues, or relying on strong assumptions. To address these limitations and better fit our real-world scenario, pre-rank in Shopee Search, we introduce in this paper ResFlow, a lightweight multi-task learning framework that enables efficient cross-task information sharing via residual connections between corresponding layers of task networks. Extensive experiments on datasets from various scenarios and modalities demonstrate its superior performance and adaptability over state-of-the-art methods. The online A/B tests in Shopee Search showcase its practical value in large-scale industrial applications, evidenced by a 1.29\% increase in OPU (order-per-user) without additional system latency. ResFlow is now fully deployed in the pre-rank module of Shopee Search. To facilitate efficient online deployment, we propose a novel offline metric Weighted Recall@K, which aligns well with our online metric OPU, addressing the longstanding online-offline metric misalignment issue. Besides, we propose to fuse scores from the multiple tasks additively when ranking items, which outperforms traditional multiplicative fusion.

\end{abstract}




% \received{8 February 2024}

\maketitle

\section{Introduction}

Modern large-scale recommender systems and search engines heavily rely on modeling diverse user feedback to understand the preferences of users and better provide personalized services. Specifically, for e-commerce platforms like Meituan~\cite{xi2021modeling}, AliExpress~\cite{li2020improving}, Taobao~\cite{ma2018entire}, Walmart~\cite{wu2022multi}, and Shopee, estimating the click-through rate (CTR) and the click-through \& conversion rate (CTCVR) of a user w.r.t. items have become one of their primary tasks. These metrics serve as their main indicators when ranking items. 

Given that these estimation tasks are closely related and that high-commitment user behaviors exhibit significant sparsity, e.g., CTCVR is typically at the level of 0.1\%, multi-task learning (MTL)~\cite{zhang2021survey,zheng2022survey} has become an integral part of such systems to boost cross-task information interchange and mitigate the sample sparsity issue. 

\begin{figure}[t]
  % \centering
  \vspace{2.5pt}
  \includegraphics[width=\linewidth]{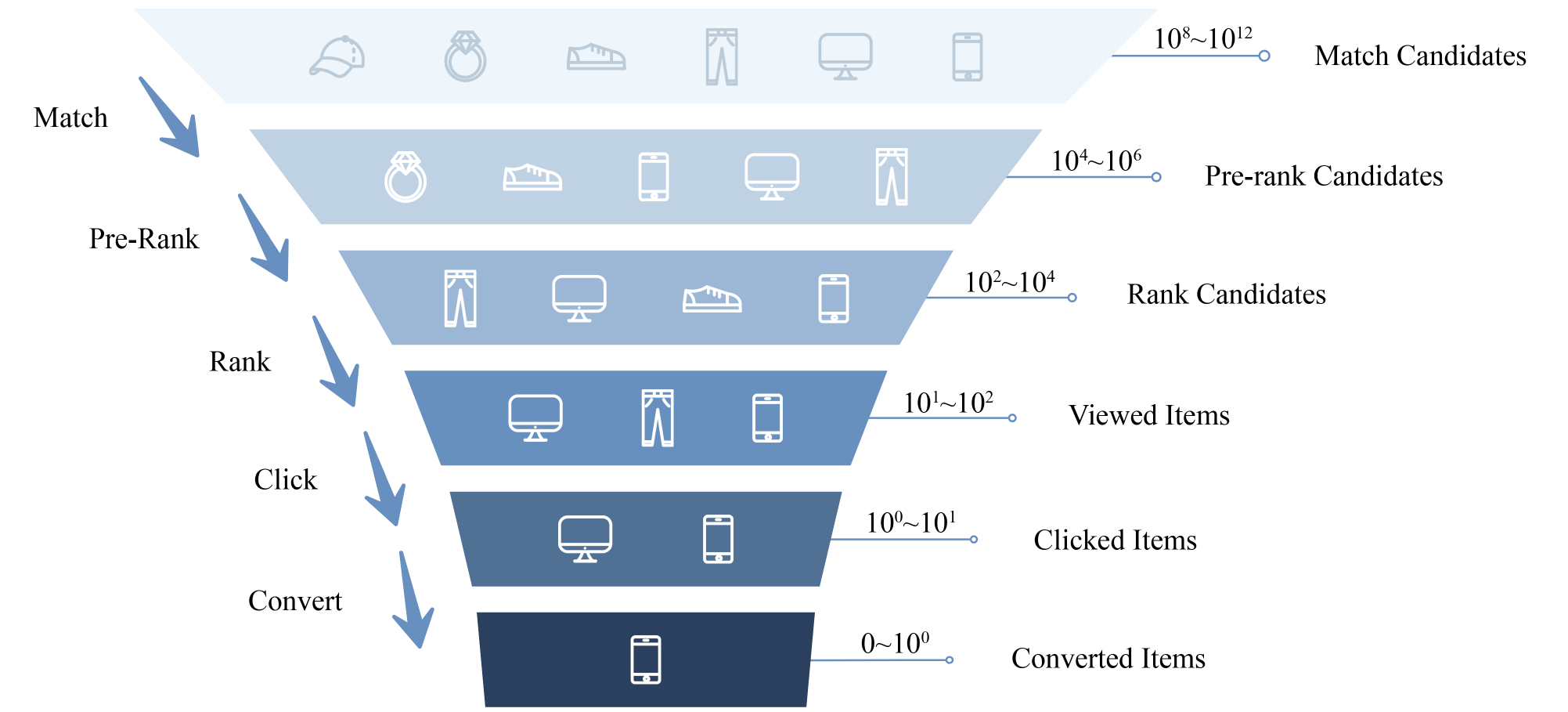}
  \caption{An illustration of the cascading item filtering process by the multi-stage candidate selection framework in large-scale applied ranking systems and subsequent user feedback. It sketches the quantities of samples at each stage, highlighting the sparsity of conversion feedback.} 
  \label{fig:multi-stage-and-sparsity}
\end{figure}

However, real-world large-scale e-commerce platforms have unique features that make MTL for their ranking systems special and challenging. Firstly, large-scale ranking systems commonly use a multi-stage candidate selection framework~\cite{zhang2023rethinking,wang2020cold}, e.g., match $\shortrightarrow$ pre-rank $\shortrightarrow$ rank, to strike a balance between efficiency and accuracy, as depicted in Figure~\ref{fig:multi-stage-and-sparsity}. In the early stages, such as pre-rank, the system must quickly sort through millions of items with limited resources, shortlisting a manageable number of candidates that align well with user interests and search keywords for the next stage. This requirement poses a significant challenge to the efficiency of model inference, forbidding the employment of complex models. Secondly, user actions in e-commerce platforms are typically designed to be carried out progressively, e.g., users can only place an order after clicking the item. Effective leverage of such sequential dependence properties may have critical importance. 

Among the mainstream multi-task learning methods, MMOE~\cite{ma2018modeling} and PLE~\cite{qin2020multitask} foster cross-task information sharing through expert-based network architectures. However, they lack explicit modeling of task relationships, showing relatively inferior performance~\cite{xi2021modeling}. Towards leveraging the sequential dependence among tasks, AITM \cite{xi2021modeling} proposed to transfer information sequentially from the former task to the latter via an attention-based module. However, its high computational intensity limits its applicability in resource-constrained scenarios like pre-rank. 

On the other hand, ESMM~\cite{ma2018entire} proposed to use causal graphs to model sequentially dependent tasks, modeling the conditional probability of the latter step given the former. Thereafter, given the inherent sample selection bias~\cite{chen2023bias} of modeling condition probability, subsequent methods including ESCM$^2$~\cite{wang2022escm2} and DCMT~\cite{zhu2023dcmt} further employed counterfactual regularizers~\cite{schnabel2016recommendations} to mitigate this issue. However, they encounter scalability issues when extending beyond two tasks, due to the high variance associated with counterfactual regularizers and their accumulation along the causal dependency chain~\cite{wang2022escm2}. Besides, the applicability of these causal methods is limited to scenarios where clear causal relationships exist. 

To address the aforementioned limitations and better fit our real-world scenario, we propose \textit{ResFlow}, a lightweight multi-task learning framework that boosts efficient information transfer from one task to another by introducing a set of residual connections between corresponding layers of their networks. ResFlow is hence generally applicable in situations where the information from the former task is beneficial to the latter~\cite{wu2022multi,xi2021modeling,tao2023task,jin2022multi}, typically from low-commitment, dense tasks to high-commitment, sparser ones, including sequentially dependent cases, e.g., "click" $\shortrightarrow$ "order", and more general ones where tasks show certain progressiveness, e.g., "like" and "forward". It is worth noting that such residual connections can be straightforwardly extended to longer progressive chains, e.g., "click" $\shortrightarrow$ "add-to-cart" $\shortrightarrow$ "order", fostering a sufficient and continuous flow of information, as illustrated in Figure \ref{fig:ResFlow}. With its simplicity and generality, ResFlow can be integrated smoothly into diverse ranking stages and application scenarios. 

Comprehensive experiments on various offline datasets, as well as online A/B tests in Shopee Search, have validated the superb effectiveness and scalability of ResFlow. In particular, according to our online A/B tests, ResFlow brings a 1.29\% increase in OPU (order-per-user) without extra system latency. ResFlow is now fully deployed in the pre-rank module of Shopee Search.

Furthermore, to facilitate efficient online deployment, we propose a new offline metric Weighted Recall@K, which aligns well with our online metric OPU, addressing the longstanding issue of online-offline metric misalignment. Besides, we propose to fuse the scores from the multiple tasks additively when ranking items, which according to our experiments consistently outperforms traditional multiplicative score fusion.

The contribution of this paper can be summarized as:
\begin{itemize}[leftmargin=*,topsep=1pt]
    \item We propose ResFlow, a novel, lightweight, and versatile MTL framework, demonstrating its superior performance and adaptability over state-of-the-art methods and fully deploying it in the pre-rank module of Shopee Search. 
    \item We propose a new offline metric that addresses the longstanding online-offline metric alignment issues in practical deployment. We propose to additively fuse multi-task scores when ranking items, which outperforms traditional multiplicative score fusion.
    % Our investigation of the practical deployment of ResFlow also provides insights into key implementation challenges of employing multi-task learning in an applied ranking system, including online-offline metric alignment and multi-score fusion strategies. 
\end{itemize}

\section{Related Works} \label{sec:related}

Multi-task learning, originating from Caruana's work~\cite{caruana1997multitask}, has evolved through various techniques~\cite{misra2016cross,zhang2010probabilistic,yang2017trace,zhang2017learning}. Its application in e-commerce has recently seen a significant increase, which can be mainly categorized into: 1) parameter-sharing based~\cite{ma2018modeling,qin2020multitask,du2022hierarchical}; 2) feature information transfer based~\cite{xi2021modeling,jin2022multi,tao2023task,wu2022multi}; and 3) probability transfer based~\cite{ma2018entire,wang2022escm2,zhu2023dcmt}. ResFlow aligns with the feature information transfer category. % Unlike some studies that pursue Pareto optimality among multiple task objectives~\cite{lin2019pareto,momma2022multi,xie2021personalized,guangyuan2022recon}, ResFlow adopts a single joint objective, a mainstream choice in MTL. 

% We involve a single joint objective for multi-task learning which is the mainstream choice, while some others seek Pareto optimality among the multiple objectives from each task \cite{lin2019pareto,momma2022multi,xie2021personalized,yu2020gradient,liu2021conflict,guangyuan2022recon}. 

Our online tests are conducted within the pre-rank module of a multi-stage candidate selection framework~\cite{zhang2023rethinking,wang2020cold}. Specific stage focused works include: DeepMatch~\cite{huang2013learning} and DSSM~\cite{lu2013deep} for match stage, \citet{wang2020cold} and \citet{zhang2023rethinking} for pre-rank stage. Joint optimization across different stages has also been explored \cite{gallagher2019joint,qin2022rankflow}. 

% Our online experiments are conducted with the pre-rank module of the multi-stage candidate selection framework \cite{zhang2023rethinking,wang2020cold}. Works that focus on a certain stage of such system include DeepMatch \cite{huang2013learning} and DSSM \cite{lu2013deep}, which focused on the matching stage, and \citet{wang2020cold} and \citet{zhang2023rethinking} that also focused on the pre-rank stage. Besides, joint optimization of ranking models across different stages has been explored by \citet{gallagher2019joint} and \citet{qin2022rankflow}. 

ResFlow adopted the residual learning concept, a principle established by ResNet~\cite{he2016deep}. Various studies~\cite{lee2020residual,rebuffi2017learning} have investigated its effectiveness, and it remains popular in diverse applications, such as ControlNet~\cite{zhang2023adding} in diffusion model control and LoRA~\cite{yu2023low} in large language model adaptation. 

Some tasks in applied ranking, e.g., the CTCVR estimation, confront severe sample imbalance issues. Besides employing multi-task learning, among the typical single-task solutions, such as sampling-based \cite{batista2004study}, cost-sensitive \cite{chawla2004special,maloof2003learning}, and kernel-based \cite{wu2004aligning}, we further adopt the sampling-based, implemented as sample re-weighting, for its simplicity and compatibility with neural networks.

\section{Methodology} \label{sec:method}

\begin{figure*}
  \centering
  \includegraphics[width=\linewidth]{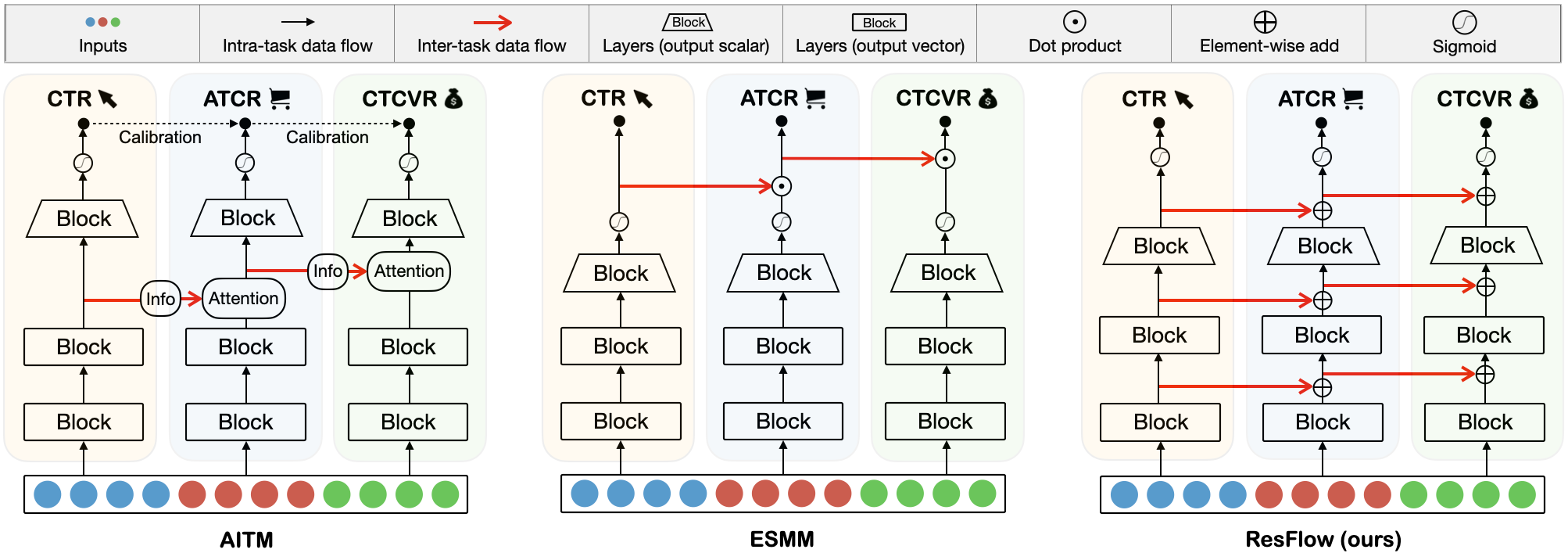}
  \caption{The conceptual architectures of ResFlow and its main competitors. Illustrated with three sequentially dependent tasks: CTR, the click-through rate; ATCR, the post-view add-to-cart rate; CTCVR, the click-through \& conversion rate. AITM extracts information from the last feature layer of the former task, transferring it to the latter via an attention-based module. ESMM models the (conditional) probability of each step, estimating ATCR and CTCVR based on the products of these probabilities. ResFlow builds residual connections between corresponding layers among tasks, enabling information transfer sufficiently at various levels in an additive manner, which is not only extremely lightweight but also shows distinguished effectiveness.} 
  \label{fig:ResFlow}
\end{figure*} 

In this section, we introduce the proposed multi-task learning framework ResFlow. As illustrated in Figure \ref{fig:ResFlow}, it boosts efficient information transfer from one task to another by introducing residual connections between corresponding layers of their networks at various levels. ResFlow is generally applicable if only the information from the former task is beneficial to the latter. 

Typical application scenarios include sequentially dependent tasks, e.g., "click" $\shortrightarrow$ "add-to-cart" $\shortrightarrow$ "order", and tasks that show certain progressiveness, e.g., "view", "like" and "forward". Note that, for information boost, it is typically connected from low-commitment, dense tasks to high-commitment, sparser ones (see Section \ref{sec:general-multi-task}).

\subsection{Multi-Task Optimization}

Given a set of related tasks, multi-task learning typically optimizes a joint loss that is a weighted sum of the loss from each task: 
\begin{equation}
    L(\vartheta,\theta_1,...,\theta_K) = \sum_{k=1}^{K} \omega_k L_k(\vartheta,\theta_k),
\end{equation}
where $K$ denotes the number of tasks, $\omega_k$ and $\theta_k$ denote the loss weight and the task-specific parameters of task $k$ respectively, $\vartheta$ denote the possible parameters that shared across tasks, and $L_k$ denotes the loss of task $k$. More specifically,  
\begin{equation} 
    L_k(\vartheta,\theta_k) = \sum_{i} \delta_k(y^i_k, \hat{y}^i_k(\vartheta,\theta_k,x^i_k)), 
\end{equation}
where $\delta_k$ denotes the loss function, e.g., cross-entropy, of task $k$, $y^i_k$ and $\hat{y}^i_k$ respectively denote the target and the prediction of $x^i_k$, the $i$-th sample of task $k$. 

\subsection{ResFlow Architecture} \label{sec:ResFlow-arch} 

The core element of ResFlow is the residual connections between corresponding layers among tasks, which are also the only extra things that need to be introduced when employing ResFlow. 

The residual connection here is a generalized one, which links a position in one task network to a position in another task network, diverging from the traditional pattern, e.g., in \citet{he2016deep} where such connections are within a single network. It allows for information transfer from the former position to the latter through element-wise addition of their values. 

Notably, residual connections are directional, it is hence required to ascertain the task relationships and determine how the tasks should be chained beforehand. Regardless, for sequentially dependent tasks, we just need to follow its dependence chain. More general cases will be studied in the experiments (Section \ref{sec:general-multi-task}). 

Besides, a logically reasonable residual connection would require its two endpoints to have a certain correspondence. ResFlow hence requires that the task networks to be residually connected have an analogous structural topology, to facilitate sensible constructions of residual connections between their corresponding points that share the same functionality and dimensionality. 

For clarity, we here formally discuss the setting where the networks for each task are of the same sequential (i.e., with no branch) architecture, e.g., the same multilayer perceptron (MLP), while illustrating some more complex settings in Appendix~\ref{sec:app-a}. 

Assume that each task network consists of $L$ sequentially chained function blocks, denoted by $f^l_k$ with $1 \le l \le L$ and $1 \le k \le K$ respectively. A function block can simply be a linear transformation followed by a nonlinear activation, while more complex structures are equally applicable. In particular, the output of the last function block would be of the target dimension, typically a scalar.  

Denoting the output of the $l$-th block (after residual, if has) of the $k$-th task as $o^l_k$, if the $(k\!-\!1)$-th task is residually connected with the $k$-th task  and there is a link behind the $l$-th function block, then: 
\begin{equation}
    o^l_k = o^l_{k-1} + f^l_k(o^{l-1}_{k}), 
\end{equation}
otherwise, it will simply be:
\begin{equation}
    o^l_k = f^l_k(o^{l-1}_{k}),
\end{equation}
where $o^{0}_{k}$ refers to the input of task $k$.

Essentially, having a residual link behind would redefine $f^l_k$ as a residual learner that learns to yield $o^l_k$ additively based on the output $o^l_{k-1}$ from the former task. Making use of the feature learned by the former task, which typically has denser positive feedback, could make the latter's learning easier. Otherwise, it needs to learn fully by itself to generate $o^l_k$. 

To seek a maximum information transfer, such residual links can be employed behind all function blocks of chained task networks, including the final ones that output the logits, as in Figure~\ref{fig:ResFlow}. According to our experiments, all of these connections are beneficial. 

With these residual connections, ResFlow enables transferring information effectively and sufficiently at various levels, leading to improved performance. The additional residual connections come with minimal extra computational overhead and thus can be used in various scenarios, including those with limited computational resources like pre-rank. Without leveraging any harsh prior knowledge, it is generally applicable for joint learning of tasks where the information from one task benefits another. % some other(s). % its subsequent task. 

\subsection{Implementation Details}

Although ResFlow can be generally applicable, task definitions can still be important. For example, one can choose to build a task to directly model CTCVR, but given there is an auxiliary CTR task, one can also choose to model CTCVR indirectly via modeling its probability condition on the estimated CTR, like ESMM \cite{ma2018entire}. According to our experiments, with ResFlow, direct modeling leads to better performance than modeling the conditional. Apart from the sample selection bias issue in modeling conditional probability \cite{chen2023bias,wang2022escm2,zhu2023dcmt}, direct modeling of CTCVR could offer better progressiveness upon the CTR task, making residual connections in-between more acceptable. From another perspective, residual learning could play the role of additively encoding in the conditional probabilities. % better fit with the residual learning framework 

% For example, in choosing a task to residual CTR estimation, we empirically find using CTCVR better than CVR like ESMM \cite{ma2018entire}. This could be attributed to the larger gap between CTR and CVR estimations. CVR, defined as the post-click conversion rate, attempts to isolate the influence of click actions. Res-links between CTR and CVR force functional blocks to learn residual between tasks sharing limited information. In contrast, estimating "add-to-cart" rate in-between CTR and CTCVR significantly improves performance. These observations shed insights that tasks should be chained according to progressiveness and that we should be mindful of managing gaps between adjacent tasks. 

For sequentially dependent actions, when modeling the probabilities of progressing from the outset to each of the later stages, their predicted probability should be strictly non-increasing by definition. According to our experiments, in most e-commerce scenarios, we need no extra regularization to ensure this. They will be naturally satisfied after training, possibly due to the ground-truth probabilities having very sharp decreases within each step. Nevertheless, we tested a set of regularizers for this purpose. Among them, forcing the residual logit, i.e., the output of final block $f^L_k(o^{L-1}_{k})$, to be non-positive by taking a min between it and zero, gives the best performance, which is very stable, shows no harm to the performance, and can bring improvements in some cases.

\section{Offline Evaluation} \label{sec:offline}

This section presents experiments conducted on datasets from diverse scenarios and modalities to demonstrate ResFlow's superior performance and distinguished adaptability over existing methods. 

\subsection{Experiment Setup}

\subsubsection{Datasets}

The datasets considered include: 1) AliCCP, from real-world traffic logs of the recommender system in Taobao \cite{ma2018entire}, and its scenario-splits S0, S1, S0\&S1; 2) AE, from real-world traffic logs of the search system in AliExpress \cite{li2020improving}, comprising country-wise subsets AE-RU, AE-ES, AE-FR, AE-NL, and AE-US; 3) Shopee, a private dataset from traffic logs of the primary search scenario in Shopee, collected within 10 consecutive days; 4) KuaiRand-Pure-S1, the main scenario of the multi-scenario user interaction dataset collected from the primary recommender system of Kuaishou \cite{gao2022kuairand}; 5) MovieLens-1M, a widely used movie ratings dataset. 
Their statistics are provided in Table \ref{tab:ecommerce-dataset}, \ref{tab:kuairand-dataset}, and \ref{tab:movielen-dataset}.

\begin{table}
    \caption{Statistics of used e-commerce datasets. Vew, Clk, and Cnv denote the number of viewed, clicked, and converted samples, respectively.}
    \begin{tabularx}{0.46\textwidth}{
        >{\centering\arraybackslash}m{3em}
        >{\centering\arraybackslash}m{1.8em}
        >{\centering\arraybackslash}m{1.8em}
        >{\centering\arraybackslash}m{1.8em}
        >{\centering\arraybackslash}m{1.8em}
        >{\centering\arraybackslash}m{1.8em}
        >{\centering\arraybackslash}m{1.8em}
        >{\centering\arraybackslash}m{3.6em}
    }
    \hline
    \hline
    Dataset & Split & Vew & Clk & Cnv & CTR & CVR & CTCVR \\
    \hline
     \multirow{2}{*}{AliCCP}& train & $42M$ & $1.6M$ & $9.0K$ & $3.9\%$ & $0.55\%$ & $0.021\%$ \\
     & test & $43M$ & $1.7M$ & $9.4K$ & $3.9\%$ & $0.56\%$ & $0.022\%$ \\
     \hline
     \multirow{2}{*}{S0} & train & $26M$ &$1.0M$ & $5.7K$ & $3.8\%$ & $0.57\%$ & $0.021\%$\\ 
    & test & $26M$ & $1.0M$ & $6.0K$ & $3.8\%$ & $0.60\%$ & $0.023\%$ \\
    \hline
     \multirow{2}{*}{S1} & train & $16M$ & $0.6M$ & $3.3K$ & $4.0\%$& $0.52\%$ & $0.021\%$ \\
     & test & $16M$ & $0.7M$ & $3.4K$ & $4.0\%$ & $0.52\%$ & $0.021\%$ \\
     \hline
    \multirow{2}{*}{S2} & train & $0.3M$ & $14K$  & $0$ & $4.6\%$  & $0\%$ & $0\%$\\
    & test & $0.3M$ & $14K$ & $0$ & $4.6\%$ & $0\%$ & $0\%$ \\
     \hline
    \multirow{2}{*}{S0\&S1} & train & $42M$ & $1.6M$ & $9.0K$ & $3.9\%$ & $0.55\%$ & $0.021\%$\\
    & test & $43M$ & $1.7M$ & $9.4K$ & $3.9\%$ & $0.56\%$ & $0.022\%$ \\
    \hline
    \hline
     \multirow{2}{*}{AE-ES} & train & $22M$ & $0.6M$ & $13K$ & $2.6\%$ & $2.25\%$ & $0.058\%$\\
     & test & $9.3M$ & $0.3M$ & $6.1K$ & $2.8\%$ & $2.30\%$ & $0.066\%$\\
     \hline
     \multirow{2}{*}{AE-FR} & train & $18M$ & $0.34M$ & $9.0K$ & $1.9\%$ & $2.63\% $& $0.049\%$ \\
     & test & $8.8M$ & $0.2M$ & $5.3K$ & $2.2\%$ & $2.71\%$ & $0.061\%$\\
     \hline
     \multirow{2}{*}{AE-NL} & train & $12.2M$ & $0.25M$ & $8.9K$ & $2.0\%$ & $3.63\%$ & $0.073\%$ \\
     & test & $5.6M$ & $0.14M$ & $4.9K$ & $2.4\%$ & $3.61\%$ & $0.088\%$\\
      \hline
    \multirow{2}{*}{AE-US} & train & $20.0M$ & $0.29M$ & $7.0K$ & $1.5\%$ & $2.41\%$ & $0.035\%$\\
     & test & $7.5M$ & $0.16M$ & $3.9K$ & $2.2\%$ & $2.41\%$ & $0.052\%$\\
      \hline
    \multirow{2}{*}{AE-RU} & train & $95M$ & $2.6M$ & $43K$ & $2.7\%$ & $1.68\%$ & $0.045\%$\\
     & test & $35M$ & $1.0M$ & $19K$ & $3.0\%$ & $1.79\%$ & $0.054\%$\\
    \hline
    \hline
    \multirow{2}{*}{Shopee} & train & $4.1B$ & $348M$ & $8.9M$ & $8.4\%$ & $2.57\%$ & $0.217
    \%$\\
    & test & $0.48B$ & $40M$ & $1.0M$ & $8.3\%$ & $2.65\%$ & $0.220\%$\\
    \hline
    \hline
    \end{tabularx}
    \label{tab:ecommerce-dataset}
\end{table}

\begin{table}
    \caption{Statistics of the KuaiRand-Pure-S1 dataset. We list the amount of different user feedback. VV denotes Valid View, defined by the data provider, indicating high confidence in actual user engagement.}
    \begin{tabular}{ccccccc}
    \hline
    \hline
     Split & Total & VV & Like & Follow & Comment & Forward \\
    \hline
     train & $945K$& $500K$ & $21K$ & $1.1K$ & $2.8K$ & $1.1K$ \\
     test & $102K$ & $50K$ & $2K$ & $130$ & $318$ & $107$ \\
     \hline
    \hline
    \end{tabular}
    \label{tab:kuairand-dataset}
\end{table}

\begin{table}
    \caption{Statistics of the MovieLens-1M dataset. We list the number of samples with different ratings.}
    \begin{tabular}{ccccccc}
    \hline
    \hline
    Split & Total & 1 & 2 & 3 & 4 & 5 \\
    \hline
    train & $0.80M$ & $45.3K$ & $84.6K$ & $0.207M$ & $0.278M$ & $0.185M$ \\
    test &$0.20M$ & $10.8K$ & $22.9K$ & $54.0K$ & $70.6K$ & $41.4K$  \\
     \hline
    \hline
    \end{tabular}
    \label{tab:movielen-dataset}
\end{table}
In particular, the Shopee dataset has three types of labels: click, add-to-cart, and conversion. Shopee-2 indicates results on using only the click and conversion labels, i.e., there are two tasks, CTR and CTCVR estimation, while Shopee-3 indicates results using all three labels with three estimation tasks, CTR, ATCR, and CTCVR. 

KuaiRand-Pure-S1 and MovieLens-1M are used to attest ResFlow's applicability to more general scenarios and other modalities. 

\subsubsection{Baseline Methods}

The baseline methods considered in our experiments include: ESMM \cite{ma2018entire}; AITM \cite{xi2021modeling}; ESCM$^2$-IPW \cite{wang2022escm2}; ESCM$^2$-DR \cite{wang2022escm2}; DCMT \cite{zhu2023dcmt}. Besides, we include NSE (naive shared embedding), which employs a separate network for each task but shares the embedding table across tasks. We exclude multi-task methods such as MMOE \cite{ma2018modeling} and PLE \cite{tang2020progressive} due to their relatively inferior performance as reported in baselines \cite{xi2021modeling,wang2022escm2,zhu2023dcmt}. 

\subsubsection{Evaluation Metrics}

In line with the convention \cite{ma2018entire,wang2022escm2,zhu2023dcmt,chen2023clustered}, we employ the area under the ROC curve (AUC) as the main evaluation metric for ranking tasks, while adopting the mean square error (MSE) for regression tasks. All experiments are conducted five times with different random seeds, and the average value along with the standard deviation is provided. 

\subsubsection{Implementations}

To ensure a fair comparison, we use the same backbone network architecture (which we mainly considered MLP with two or three hidden layers) and the same basic configurations, e.g., embedding size, optimizer, and learning rate, for all tasks and all methods in our experiments. Besides, we tune the optimum sample re-weighting ratio for each task for all methods. 

% We mainly considered MLP with two or three hidden layers as the backbone network architecture, while for the Shopee dataset, we adopt a twin-tower architecture, one for user and query and one for item, which can be in favor of direct deployment in pre-rank 

Note that owing to the particular nature of our considered multi-task settings, e.g., CTR and CTCVR are predicting the rates of the same user w.r.t. the same item, the inputs of different task networks in our experiments are typically the same and shared. 

% Besides, in our implementation, all continuous features are discretized and then used in the same way as ID/category features, fetching their vectorized representation from the embedding table. 

Due to page limitations, we provide more detailed information about the datasets and implementations in Appendix~B. 

\subsection{Performance on E-commerce Datasets}
\label{sec:exp-e-commerce-perf}

\begin{table*}[ht]
    % \centering
    \caption{The AUC results of the CTCVR estimation task on offline e-commerce datasets. The best results are presented in bold font, while the second bests are marked with underlines. ResFlow consistently achieves the best.} 
    % "/" indicates no result was produced for some reason.  
    %"*" denotes the best-known result of the corresponding method reported in prior works.
    % "--" means no record is found.
    \begin{tabularx}{1\textwidth}{
        >{\centering\arraybackslash}m{3.8em}
        >{\centering\arraybackslash}m{4.8em}
        >{\centering\arraybackslash}m{4.8em}
        >{\centering\arraybackslash}m{5.2em}
        >{\centering\arraybackslash}m{5.2em}
        >{\centering\arraybackslash}m{5.4em}
        >{\centering\arraybackslash}m{4.8em}
        >{\centering\arraybackslash}m{5.2em}
        >{\centering\arraybackslash}m{6.4em}
    }
    \hline
    \hline
    Dataset & NSE & PLE & AITM & ESMM & ESCM$^2$-IPW & ESCM$^2$-DR & DCMT & ResFlow (ours) \\
    \hline
    S0 & $0.619\pm0.001 $ & $0.630 \pm 0.005$ & $0.636\pm0.002$ & $0.632\pm0.001$ & $0.624\pm0.007$ & $0.622\pm0.006$ & $\underline{0.640\pm0.001}$ & $\mathbf{0.656\pm0.001}$ \\
    
    S1 & $0.621\pm0.002$ & $0.624 \pm 0.003$ & $0.623\pm0.001$ & $0.623\pm0.003$ & $0.624\pm0.002$ & $0.623\pm0.003$ & $\underline{0.632\pm0.002}$ & $\mathbf{0.647\pm0.002}$ \\
    
    S0\&S1 & $0.623\pm0.002$ & $0.635 \pm 0.001$ & $0.637\pm0.001$ & $0.637\pm0.003$ & $0.623\pm0.005$ & $0.623\pm0.001$ & $\underline{0.634\pm0.002}$ & $\mathbf{0.661\pm0.001}$ \\
    
    AliCCP & $0.624\pm0.002$ & $0.640 \pm 0.002$ & $\underline{0.644\pm0.001}$ & $0.641\pm0.003$ & $0.625\pm0.005$ & $0.627\pm0.003$ & $0.643\pm0.002$ & $\mathbf{0.664\pm 0.002}$\\
    
    \hline
    AE-ES & $0.861\pm0.002$ & $0.873 \pm 0.002$& $0.876\pm0.001$ & $0.867\pm0.003$ & $0.868\pm0.003$ & $0.871\pm0.002$ & $\underline{0.886\pm0.003}$ & $\mathbf{0.893\pm0.001}$ \\
    
    AE-FR & $0.842\pm0.002$ & $0.852 \pm 0.002$ & $0.856\pm0.002$ & $0.851\pm0.001$ & $0.873\pm0.007$ & $0.870\pm0.005$ & $\underline{0.874\pm0.003}$ & $\mathbf{0.885\pm0.001}$ \\
    
    AE-NL & $0.831\pm0.003$ & $0.844 \pm 0.002$ & $0.843\pm0.001$ & $0.840\pm0.002$ & $0.854\pm0.006$ & $0.854\pm0.004$ & $\underline{0.858\pm0.001}$ & $\mathbf{0.864\pm0.001}$ \\
    
    AE-US & $0.826\pm0.001$ & $0.851 \pm 0.003$ & $0.843\pm0.002$ & $0.827\pm0.002$ & $0.843\pm0.006$ & $0.841\pm0.003$ & $\underline{0.863\pm0.003}$ & $\mathbf{0.872\pm0.001}$ \\
    
    AE-RU & $0.870\pm0.003$ & $0.886 \pm 0.001$ & $0.882\pm0.002$ & $0.878\pm0.002$ & $0.882\pm0.006$ & $0.880\pm0.006$ & $\underline{0.887\pm0.001}$ & $\mathbf{0.913\pm0.002}$ \\

    \hline
    Shopee-2 & $0.865\pm0.001$ & $0.860 \pm 0.003$ & $0.855\pm0.006$ & $\underline{0.882\pm0.001}$ & $0.840\pm0.012$ & $0.812\pm0.041$ & $0.867\pm0.001$ & $\mathbf{0.902\pm0.001}$ \\
    
    Shopee-3 & $0.877\pm0.002$ & $0.877 \pm 0.003$ & $0.871\pm0.003$  & $\underline{0.893\pm0.001}$ & / & / & / & $\mathbf{0.910\pm0.002}$ \\
    % \hline
    % AliCCP$^*$ & 0.6003\cite{wang2022escm2} & $0.6532\pm0.0049$\cite{ma2018entire} &$0.6525\pm0.0024$\cite{xi2021modeling} & 0.6189\cite{wang2022escm2} & 0.6245\cite{wang2022escm2} & 0.6341 \cite{zhu2023dcmt} & --\\
    
    % AE-ES$^*$ & -- & $0.6027$\cite{zhu2023dcmt} &$0.7044$\cite{zhu2023dcmt} & 0.8688\cite{zhu2023dcmt} & 0.8681\cite{zhu2023dcmt} & 0.8817 \cite{zhu2023dcmt} & --\\
    
    % AE-FR$^*$ & -- & $0.6039$\cite{zhu2023dcmt} &$0.6923$\cite{zhu2023dcmt} & 0.8647\cite{zhu2023dcmt} & 0.8668\cite{zhu2023dcmt} & 0.8765 \cite{zhu2023dcmt} & --\\
    
    % AE-NL$^*$ & -- & $0.5329$\cite{zhu2023dcmt} &$0.5031$\cite{zhu2023dcmt} & 0.8486\cite{zhu2023dcmt} & 0.7196\cite{zhu2023dcmt} & 0.8513 \cite{zhu2023dcmt} & --\\
    
    % AE-US$^*$ & -- & $0.5881$\cite{zhu2023dcmt} &$0.5447$\cite{zhu2023dcmt} & 0.8385\cite{zhu2023dcmt} & 0.7625\cite{zhu2023dcmt} & 0.8629 \cite{zhu2023dcmt} & --\\
    \hline
    \hline
    \end{tabularx}
    \label{tab:performance comparison}
\end{table*}

We present in Table~\ref{tab:performance comparison} the experiment results on the e-commerce datasets in terms of the AUC of the CTCVR estimation task, the primary goal. As we can see, ResFlow consistently outperforms all the baselines on all datasets, achieving an average improvement of $1.54\%$ in CTCVR AUC relative to the best-performing baselines. The results for other tasks, e.g., CTR estimation, are provided in Appendix~B, where ResFlow also gets closely the best performances. 

% On the nine public datasets (excluding Shopee), ResFlow demonstrates a larger AUC uplift on datasets with smaller overall CVR (indicated in Table~\ref{tab:dataset}). This can be attributed to ResFlow's efficient and multi-abstraction-level information transfer mechanism. Smaller CVR values indicate more conflicts (clicked but no conversion) between click and conversion estimation tasks. Allowing the model to learn residuals among tasks with various sparsity helps it benefit from task differences rather than being hindered by conflicts. 

% On the large-scale industry dataset Shopee, ResFlow achieves a 2\% increment in CTCVR AUC. This improvement is likely due to public datasets representing a small subset of daily training data in the industry. With more data, ResFlow can learn more effectively than other methods. 

% The reproduced results are generally better or at least on par with the best-reported results in prior works. This is because we apply positive sample re-weighting to all baselines and tune them for optimal performance. All baselines benefit from this technique. However, positive sample re-weighting does not show a significant impact on the Shopee dataset, mainly due to insufficient training. 

% All methods achieve much higher AUC on AE and Shopee datasets than on AliCCP. This is because, with more artificial features, the estimation task becomes easier. Directly learning from original user behavior logs remains a challenging task. 

Notably, ESMM achieves the best results on the Shopee datasets among the baselines, which was the method deployed online prior to ResFlow. NSE also works reasonably well for CTCVR estimation, which may have suggested that embedding sharing along with sample re-weighting can fairly counter the sample sparsity issue. 

Although ESCM$^2$ tends to have better performance on the AE datasets, it performs less satisfactorily on the Shopee and AliCPP datasets. Such instability in performance may be attributed to the numerical sensitivity of its causal debiasing modules (IPW and DR). The performance of DCMT is relatively much better and more stable, which may be attributed to its more advanced debiasing technique. However, it still performs worse than ESMM on the Shopee dataset. % explanations here. why Shopee is special? why ESMM is better. 
Besides, according to our experiments when applied to three-task settings, i.e., Shopee-3, the training of all these causal debiasing methods (ESCM$^2$ and DCMT) consistently fails with numerical errors (NaN), while all other methods work well on Shopee-3 and lead to considerable improvement compared with Shopee-2.

\subsection{Ablation Studies \& Parameter Sensitivities} 

We validate the design of ResFlow by ablation studies. As shown in Table~\ref{tab:ablation}, residual connections for the feature blocks and the final logit can all lead to significant improvement in performance, while their combination leads to the largest improvement, which suggests the effectiveness of residual connections at various levels. It is worth noticing that when only involving a single residual connection, the performance gradually increases from "NSE + FR H1-only" (only for the first feature block) to "NSE + FR H2-only" (only for the second feature block) to  "NSE + LR" (only for logit), which suggests the residual connection for higher-level abstraction is more critical. 

To be more comprehensive, we have also tested adding residual connections under the ESMM framework. As shown in Table~\ref{tab:ablation}, feature residual can also improve ESMM's performance, which echoes the general effectiveness of residual connections in boosting information transfer. However, further introducing logit residual aggravates its performance, which should be because logit addition can not cooperate well with the probability (sigmoid-of-logit) multiplication (modeling conditional probability). On the other hand, we can see "NSE + LR" is better than "ESMM" and even "ESMM + FR", which implies that logit addition, i.e., residual connection for logit, is a more appropriate choice than probability multiplication.

\begin{figure*}
  % \centering
  \includegraphics[width=0.99\linewidth]{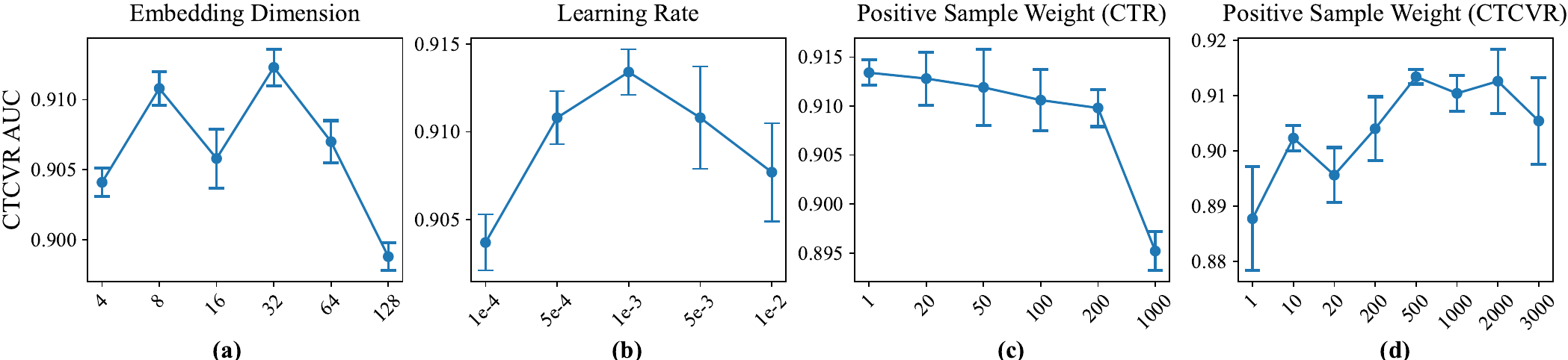}
  \caption{Parameter sensitivity experiment results. From left to right: (a) embedding dimension, (b) learning rate, (c) positive sample weight of CTR task, and (d) positive sample weight of CTCVR task. The negative sample weight is fixed as 1.} 
  \label{fig:hypter-parameters} 
\end{figure*}

We study the sensitivity of ResFlow with respect to its key parameters. The results on the AE-RU dataset are presented in Figure~\ref{fig:hypter-parameters}. We see that ResFlow is not very sensitive to its hyper-parameters. It is worth noting that the optimal positive weight for the CTR task is 1, which means it does not require sample re-weighting, though the positive samples only take around 2.7\%, being highly sparse/imbalanced in a sense. Meanwhile, for the CTCVR task, where the positive samples take around 0.045\%, we need the positive sample weight to be 500 to gain optimal performance. 

\subsection{Probing the Residual Learned by ResFlow} 

\begin{figure*}
  \includegraphics[width=1\linewidth]{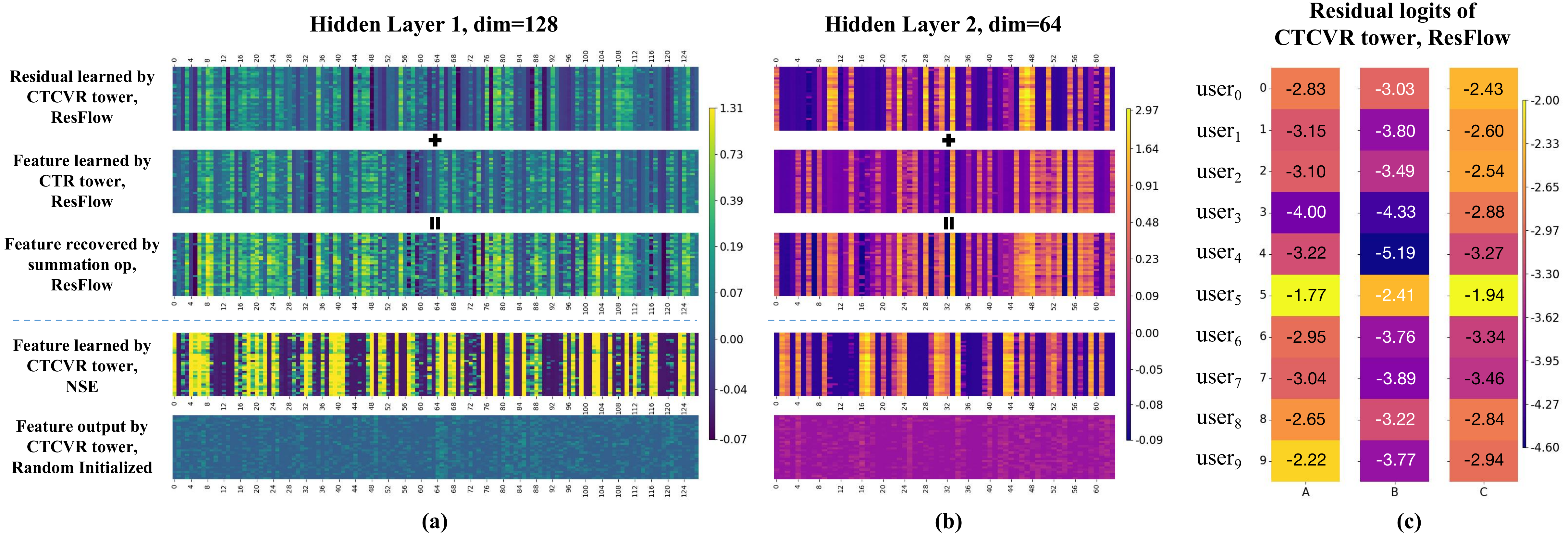}
  \caption{Case study of the learned residual. We randomly selected 10 users and sampled 3 items of different types for each user, forming 3 groups: A (click=1, order=1), B (click=1, order=0), and C (click=0, order=0). The learned residual features and logits of the CTCVR tower, and other related, are shown. Trained on the AliCCP dataset. ResFlow tends to learn less extreme features than NSE. The learned residual logits are all negative, and the values in group B are generally smaller than those in group A.} 
  \label{fig:case study} 
\end{figure*} 

To understand how ResFlow transfers information between, for example, CTR and CTCVR, in different layers, we randomly selected 10 users and sampled 3 items of different interaction types for each user forming 3 groups, where samples from group A were clicked with conversion, samples from group B were clicked without conversion, and samples from group C were not clicked. 

Figure~\ref{fig:case study}(a) and~\ref{fig:case study}(b) illustrate the hidden layer feature outputs of various modules. We can see that the residual learned by the CTCVR tower (i.e., task network) has a similar value scale as the CTR tower, resulting in the recovered (after element-wise addition) feature scale of CTCVR being relatively similar to the CTR tower. By contrast, if directly learning the CTCVR tower (with NSE), its learned feature pattern appears to be quite more extreme, primarily being either larger positive values or near-zero negative values, which may have indicated a more severe over-fitting. We further provide the feature map of a randomly initialized network for reference. 

Figure~\ref{fig:case study}(c) shows the learned residual logits by the CTCVR tower. We see that they are all negative values, and that is exactly how it should be. Because CTCVR should be strictly less than CTR, i.e., the logit should be decreased to get a smaller sigmoid probability. Notably, the learned residual logits of group B are generally smaller than group A, which is also reasonable since "click without conversion" instances require larger decreases in the logits to shift from a high probability of clicking to a low probability of conversion.

\begin{table}
    % \centering
    \caption{Ablation results of ResFlow in terms of CTCVR AUC.}
    % \resizebox{0.48\textwidth}{!}{
    \begin{tabular}{ccc}
    \hline
    \hline
    Model & AE-RU & Shopee-2 \\
    % \hline
    % M$_0$=MLP & $0.8307\pm0.0195$ & $0.8648\pm0.0012$ \\
    \hline
    NSE & $0.869\pm0.003$ & $0.865\pm0.001$ \\
    NSE + Feature Residual (FR) & $0.897\pm0.001$ & $0.891\pm0.001$ \\
    NSE + FR (H1-only) & $0.880\pm0.002$ & $0.868\pm0.002$ \\
    NSE + FR (H2-only) & $0.898\pm0.002$ & $0.887\pm0.001$ \\
    NSE + Logit Residual (LR) & $0.907\pm0.002$ & $0.896\pm0.001$ \\
    ResFlow (NSE + FR + LR) & $\mathbf{0.913\pm0.002}$ & $\mathbf{0.902\pm0.001}$ \\
    \hline
    ESMM  & $0.878\pm0.002$ & $0.882\pm0.001$ \\ 
    ESMM + FR & $ 0.899 \pm 0.004$ & $0.891 \pm 0.002$ \\ 
    ESMM + FR + LR & $0.896\pm0.006$ & $0.888\pm0.001$ \\ 
    \hline
    \hline
    \end{tabular}
    % }
    \label{tab:ablation}
\end{table}

\subsection{More General Multi-Task Scenario} 
\label{sec:general-multi-task}

% Task topology definition is important to release the potential of ResFlow fully. 

% we performed a brute-force scan on all possible topologies to figure and testify principles in determining the task topologies. 

To demonstrate the more general applicability of ResFlow. We experiment with the KuaiRand-Pure-S1 dataset, modeling its user interactions with videos, including: "valid view", "like", "follow", "comment", and "forward". 

Given that the logical relationships between "follow", "comment", and "forward" are not that clear, and that there is no significant differentiation in their frequencies, while they are all clearly less frequent than "like", and "like" is clearly less frequent than "valid view". In this experiment, we build residual connections from task "is\_valid\_view" to task "is\_like", from "is\_like" to "is\_follow", from "is\_like" to "is\_comment", and from "is\_like" to "is\_forward", respectively, and jointly train the five tasks. Although strict conditional dependence is lacking, these probability-transfer-based methods (ESMM, etc.) are also included as baselines. 

As shown in Table~\ref{tab:Kuairand-classification}, ResFlow leads to consistent improvement and achieves the best across all tasks, demonstrating its capability in modeling such types of progressiveness. Notably, AITM gets an unsatisfactory performance on the "is\_follow" task. We consider that having too many parameters in its attention-based information transfer block may have made it more likely to overfit occasionally. The full table with AUCs of all tasks is provided in Appendix~B. 

\begin{figure}
  \centering
  \includegraphics[width=0.95\linewidth]{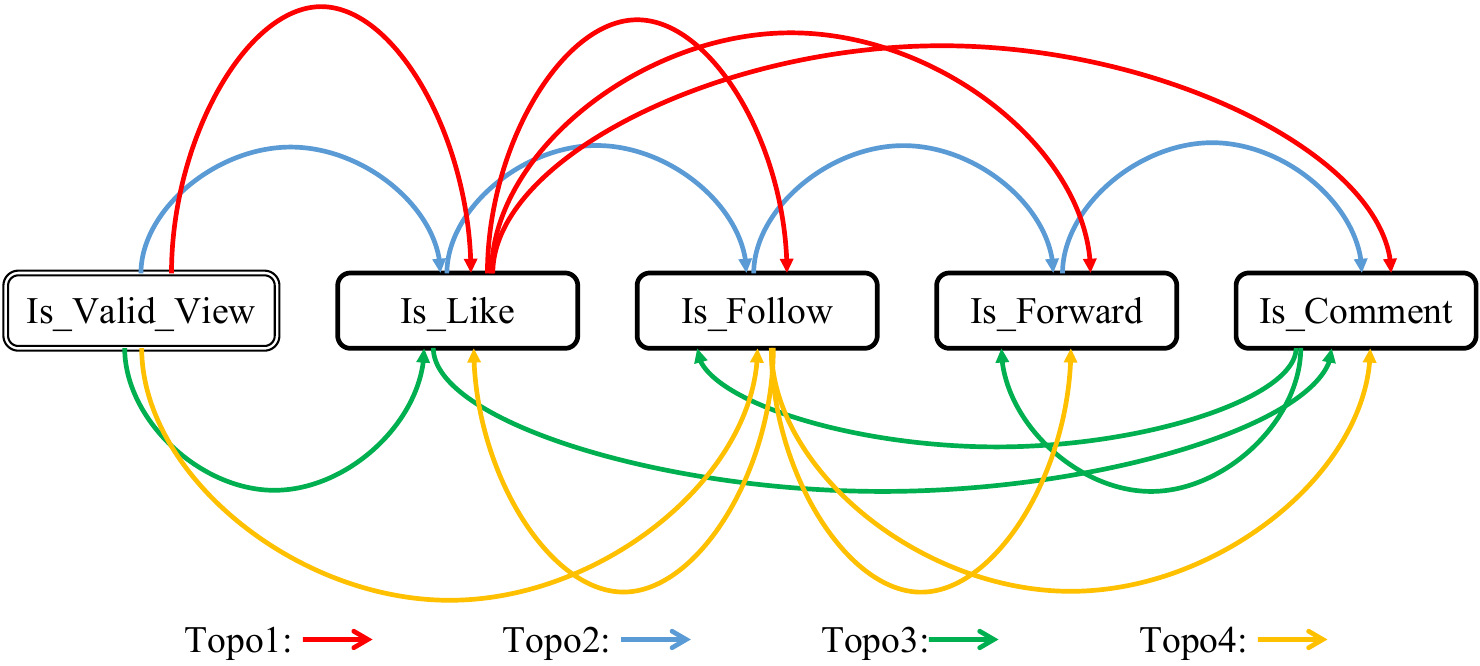}
  \caption{Visualization of different task topologies of KuaiRand-Pure-S1 Multi-Task.}
  \label{fig:topo}
\end{figure}

To further validate such an intuitive strategy, i.e., building task topologies according to their levels of sample sparsity, is a good choice, we tested a set of other possible task topologies on this dataset. We visualize them in Figure~\ref{fig:topo} and provide their respective results in Table~\ref{tab:Kuairand-topology}. We can see that "topo1", the aforementioned topology strategy, performs the best. More specifically, we see that sparse tasks can generally benefit from denser ones, e.g., task "is\_follow", "is\_forward", and "is\_comment" in topo1, and task "is\_follow" task in topo2. However, chaining tasks of the same sparsity level or putting sparse tasks before denser ones may lead to inferior performance, e.g., task "is\_forward" and "is\_comment" in topo2, task "is\_follow" in topo3, and task "is\_forward" and "is\_comment" in topo4. We also notice that task "is\_forward" seems to work properly when chained after task "is\_comment" in topo3, which may be owed to its alignment with users' overall habits on commenting and forwarding, but its accuracy is still less than the one in "topo1". This all indicates that, for scenarios without explicit dependency among tasks, building the task topology according to their sample sparsities can be a sound and easy-to-follow choice.

\begin{table}
    % \centering
    \caption{Performance on KuaiRand-Pure-S1 in terms of AUC. The bests are in bold font. The second bests are underlined.}
    \begin{tabular}{ccccc}
    \hline
    \hline
    Target & is\_follow & is\_forward & is\_comment \\
    \hline
    NSE & $0.821\pm0.005$ & $0.750\pm0.011$ & $0.779\pm0.003$ \\
    % \hline
    AITM & $0.785\pm0.014$ & \underline{$0.764\pm0.012$} & $0.782\pm0.005$ \\
    % \hline
    ESMM & \underline{$0.822\pm0.007$} & $0.757\pm0.007$ & $0.782\pm0.004$ \\    
    % \hline
    ESCM$^2$\text{-IPW} & $0.817\pm0.010$ & $0.754\pm0.008$ & $0.773\pm0.004$ \\
    % \hline
    ESCM$^2$\text{-DR} & $0.821\pm0.009$ & $0.754\pm0.008$ & $0.782\pm 0.003$ \\
    % \hline
    DCMT & $0.819\pm0.007$ & $0.762\pm0.008$ & \underline{$0.783\pm0.003$} \\
    \hline
    ResFlow & $\mathbf{0.826\pm0.002}$ & $\mathbf{0.776\pm0.006}$ & $\mathbf{0.789\pm0.003}$ \\

    \hline
    \hline
    \end{tabular}
    \label{tab:Kuairand-classification}
\end{table}

\begin{table}
    \centering
    \caption{Performance on KuaiRand-Pure-S1 in terms of AUC. The bests are in bold font. The second bests are underlined.}
    \begin{tabular}{ccccc}
    \hline
    \hline
    Target & is\_follow & is\_forward & is\_comment \\
    \hline
    ResFlow-topo1 & $\mathbf{0.826\pm0.002}$ & $\mathbf{0.776\pm0.006}$ & $\mathbf{0.789\pm0.003}$ \\
    ResFlow-topo2 & $\underline{0.825\pm0.003}$ & $0.756\pm0.010$ & $0.771\pm0.002$ \\
    ResFlow-topo3 & $0.811\pm0.007$ & $\underline{0.774\pm0.007}$ & $0.778\pm0.003$ \\
    ResFlow-topo4 & $0.820\pm0.002$ & $0.748\pm0.009$ & $0.775\pm0.002$ \\
    NSE & $0.821\pm0.005$ & $0.750\pm0.011$ & $\underline{0.779\pm0.003}$ \\
    \hline
    \hline
    \end{tabular}
    \label{tab:Kuairand-topology}
\end{table}

\subsection{Regression as Progressive Multi-Task} 
\label{sec:reg-as-progressive}

Besides predicting the likelihood of user actions, which can be formulated as classification problems, there are also practical needs to predict numerical values, e.g., predicting the video playtime or the rating of a user to an item. In this section, we showcase that such regression tasks may also be converted as progressive multi-task learning problems and effectively tackled by ResFlow. 

Formally, to predict a value that has a bounded range and indicates progressive engagements as it increases, we discretize the target variable if it is not inherently discretized. Subsequently, for each discretized threshold $v_k$, excluding the minimal one, we define a task $Q(v \!\ge\! v_k)$ to predict the likelihood that the actual value equals or surpasses $v_k$. By definition, these tasks embody progressiveness, necessitating their probabilities non-increase as $v_k$ ascends.\footnote{Forcing residual logits non-positive, taking min with 0, led to slight gains in this task.}

For setting the training targets, if the actual value is $v$, then by the definition, all tasks where $v_k \leq v$ are assigned with a label of $1$, and all other tasks receive a label of $0$. For prediction purposes, we employ the approximation of the expected value: 
% derived from the predicted probabilities: 
\[
    \mathbb{E}(v) = \sum_{k=0}^{K-1} v_k \cdot \max\big(Q(v \ge v_{k}) - Q(v \ge v_{k+1}), 0\big) + v_K \cdot Q(v \ge v_{K}), 
\]
where $v_0 < v_1 < \ldots < v_K$, $v_0$ and $v_K$ are the minimal and maximal possible values respectively, $Q(v \ge v_0) = 1$, there are $K$ tasks. 

% We force the decreasing predicted probabilities of tasks by forcing the residual to be negative. 

\begin{table}
    % \centering
    \caption{Regression performances on KuaiRand-Pure-S1 and MovieLens-1M in terms of MSE. Progressive indicates converting the regression into a progressive multi-task problem.} 
    \begin{tabular}{ccc}
    \hline
    \hline
    Dataset & KuaiRand-Pure-S1 & MovieLens-1M  \\
    \hline
    Traditional & $1719.92\pm3.82$ & $0.906\pm0.002$ \\
    % \hline 
    % Traditional-wide & $1767.57\pm3.42$ & $0.908\pm0.003$ \\
    % \hline
    Progressive + NSE & $1720.12\pm4.32$ & $0.906\pm0.002$ \\
    % \hline
    Progressive + ResFlow & $\mathbf{1658.44\pm5.21}$ & $\mathbf{0.894\pm0.002}$ \\
    \hline
    \hline
    \end{tabular}
    \label{tab:result-regression}
\end{table}

We experiment with: 1) MovieLens-1M, where we predict the users' ratings on movies; and 2) KuaiRand-Pure-S1, where we predict the users' video playtime. We compare the progressive modeling of regression with the traditional regression and test whether ResFlow can help such progressive modeling. Since multi-task solutions need to involve multiple networks and hence more parameters, we search for the best network configuration within the multilayer perception class for all methods for a fair comparison. 

As the results presented in Table~\ref{tab:result-regression}, progressive modeling (with NSE) works almost equally well as traditional regression, while ResFlow can effectively boost the performance of progressive modeling, leading to clear improvements over traditional regressions.

\section{Online Deployment} \label{sec:online}

\begin{figure*}
  % \centering
  \includegraphics[width=\linewidth]{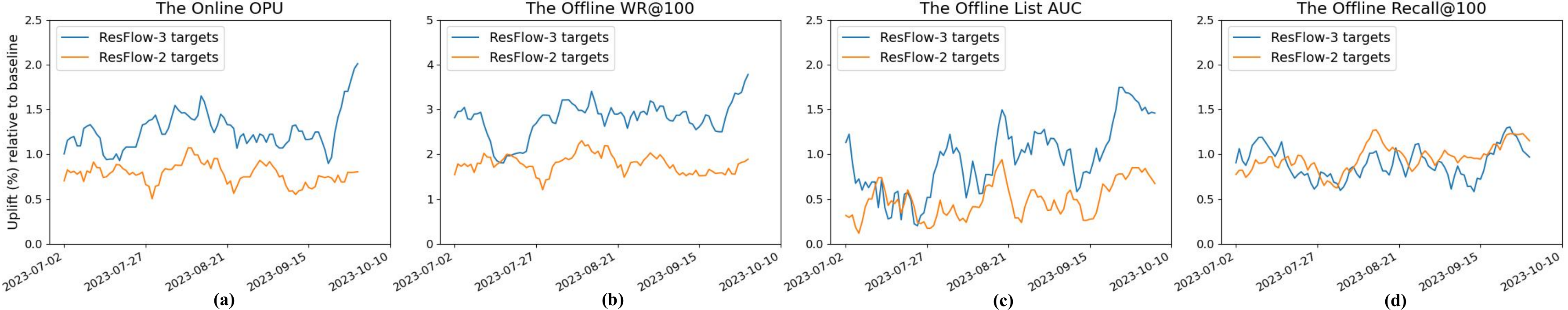}
  \caption{The uplift of online OPU and other offline metrics of ResFlow relative to the formerly deployed method (ESMM) during online A/B test. The curves are smoothed to better show the trend. Offline WR@100 aligns fairly well with online OPU. } % From left to right: (a) the online OPU; (b) the offline WR@100; (c) the offline List AUC; and (d) the offline Recall@100. % See Figure \ref{?????} in the Appendix for the non-smoothed version. % In contrast to others 
  \label{fig:online-experiment}
\end{figure*} 

% To validate ResFlow's efficiency and scalability, we conducted experiments in the pre-rank module of Shopee Search. Implementing deep neural networks like ResFlow in large-scale online services, especially in resource-sensitive stages such as pre-rank, presents several challenges: training data creation, model robustness over time, online-offline metric alignment, score fusion strategy, and latency optimization. 

We conducted prolonged A/B tests to assess the effectiveness of ResFlow in the online environment and ultimately deployed it in the pre-rank module in the primary search scenario of Shopee. In this section, we detail our investigations about the online deployment. 

\subsection{The Pre-Rank Module}

The pre-rank module is responsible for sorting roughly filtered item candidates from the match module and passes the top-ranked to the rank module. Typically, it needs to handle millions of items within a very short time, strictly less than 150ms in our scenario, hence requiring both efficiency and accuracy. 

Following the normal practice \cite{huang2013learning,wang2020cold,zhang2023rethinking}, we adopt a twin-tower task network architecture for pre-rank: one tower for extracting representation for user-and-query, and one for item, which is illustrated in Figure~\ref{fig:pre-rank-two-tower} in the Appendix. It facilitates the separation of user-query representation inference, which requires online processing, from item representations, which can be precomputed offline.

\subsection{Online-Offline Metric Alignment}

The main online metric in Shopee is OPU, the average number of orders placed by each user (in one day). OPU-driven optimization could contribute to the growth of platform scale and market share. 

% However, OPU and similar online metrics are critical for evaluating system performance but are inherently limited to online assessments. This limitation stems from the challenge of accurately simulating or predicting user responses to modifications in the ranking model offline. 
However, online OPU, as well as other similar online metrics, can only be effectively evaluated online. Regarding offline data, it is a fixed value that cannot be used to tune the model and hyper-parameters. Therefore, finding offline metrics that well aligns with the online metric is vital for efficient model optimization in the industry, given the time-intensive and costly nature of online A/B testing. However, commonly used offline metrics do not well align with the online metrics \cite{beel2013comparative,huzhang2021aliexpress,zhang2023rethinking}. Tuning configurations for online deployments hence has long been and still is a challenge. 

To address this, we propose the offline metric Weighted Recall@K (WR@K), which characterizes how well the model can rank hot items up within the top K, which we found to align fairly well with the online metric for pre-rank. Formally, 
\begin{equation}
    WR@K = \frac{\sum_{k=0}^{K} W_{k}}{\sum_{n=0}^{N} W_{n}}, 
\end{equation} 
where $K$ is the given parameter of the metric, $N$ is the total number of items to be ranked, and $W_k$ indicates the number of orders of item $k$ (in one day, under a given query). 

% For Click-Weighted Recall@K, $W_k$ indicates the number of clicks of item $k$, while for Order-Weighted Recall@K, $W_k$ indicates the number of order placements of item $k$. 

As results shown in Figure \ref{fig:online-experiment} and Table \ref{tab:result-metric-correlation}, WR@100 aligns reasonably well with OPU compared to other metrics. In the compared metrics, NDCG is the normalized discounted cumulative gain, while List AUC is the AUC of the ranked list considering items with order labels as positive. We consider that the effectiveness of WR@K stems from its incorporation of collective user feedback (i.e., $W_k$) that reflects item popularity, while metrics like List AUC, Recall@K, and NDCG tend to focus on individual user responses. 

Given that List AUC and NDCG emphasize a different perspective from WR@K and also show a positive correlation with the online metric to a certain extent, we adopt WR@K as the primary metric for evaluating model performance, while using List AUC and NDCG as complementary monitoring metrics \cite{beel2013comparative,huzhang2021aliexpress,zhang2023rethinking}. 

%, heeding the lessons from \cite{beel2013comparative,huzhang2021aliexpress,zhang2023rethinking} that caution against relying on a single offline metric. 

% We denote variations of WR@100 as WR$_{log}$@100, WR$_{sqrt}$@100, and WR$_{square}$@100, corresponding to the replacement of $W_{k}$ with $\log(W_{k})$, $W_{k}^{0.5}$, and $W_{k}^{2}$, respectively. The variants of $W_{k}$, while demonstrating a similar correlation pattern to WR@K, are slightly less effective in terms of Pearson Correlation Coefficient (PCC), mainly due to their tendency to overemphasize either long-tail orders or highly popular items. 
% by considering multiple facets of user feedback and item interaction dynamics 
% which ensures a more comprehensive and reliable assessment 

\begin{table*}
    \begin{minipage}{0.69\textwidth}
        \begin{tabular}{ccccccc}
            \hline
            \hline
            Method \& Score Formula & WR@100 & List AUC & OPU & CTR & CTCVR & BCR@20\\
            \hline
            ESMM, CTR$^{\text{0.9}}$$\times$CVR$^{\text{1.1}}$ & +0.77\% & +0.23\% & +0.56\% & -0.02\% & +0.61\% & +0.02\%\\
            ESMM, CTR*2+CTCVR*19 &+2.7\% &+0.09\%  & +0.75\% & +0.12\% & +0.67\% & -0.01\%\\
            \hline
            ResFlow, CTCVR & +1.41\% & +1.82\% & +0.45\% & -0.15\% & +0.40\% & -0.02\%\\
            ResFlow, CTR$^{\text{-0.2}}$$\times$CTCVR & +3.37\% & +1.03\% & +1.33\% & +0.06\% & +1.19\% & -0.02\%\\
            ResFlow, CTR+CTCVR*20 & +4.19\% & +0.88\% &+2.14\% & +0.66\% & +2.11\% & +0.02\%\\
            ResFlow, CTR+CTCVR*20+RL & +4.11\% & +0.82\% & +2.04\% & +0.61\% & +1.97\% & -2.13\%\\
            ResFlow, CTR+CTCVR*20+RS &+0.03\% &-0.13\% & -0.53\% & -0.19\% & -0.44\% & -2.99\%\\
            \hline
            \hline
        \end{tabular}
        \label{tab:online-exp-formula}
    \end{minipage}
    \begin{minipage}{0.3\textwidth}
        \raggedleft
        \includegraphics[width=0.9\linewidth]{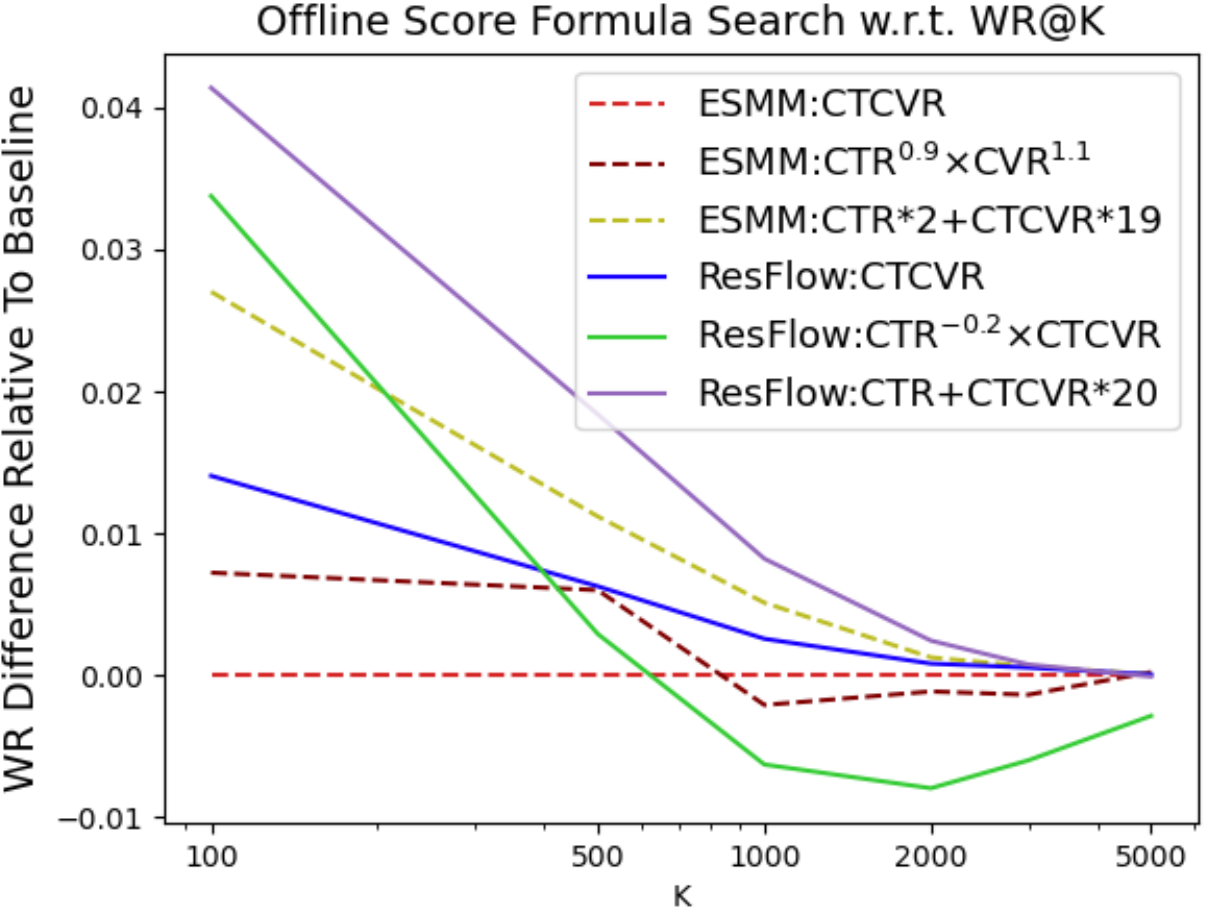}
        \label{fig:score-formula}
    \end{minipage}
    \caption{Here we use ESMM with fusion strategy CTR$^1$$\times$CVR$^1$, i.e., CTCVR, as the baseline. Left: online A/B test results on various score fusion strategies. We show the relative uplift of offline WR@100 and List AUC, and online OPU, CTR, CTCVR, and BCR@20 (bad case rate @ top 20). RL: weighted relevance level. RS: weighted relevance score. Both ESMM and ResFlow are trained with 3 targets. Right: best offline search results of score fusion formulas. We show their WR@K relative to the baseline.} 
    \label{tab:score-formula}
\end{table*}

\begin{table}
    % \centering
    \caption{Pearson correlation coefficient (PCC) between offline metric uplift and online metric uplift.}
    \begin{tabular}{c|cc|cc}
    \hline
    \hline
    \multirow{2}{*}{Metrics} & \multicolumn{2}{c|}{2 Targets Uplift} & \multicolumn{2}{c}{3 Targets Uplift} \\
    & PCC & p-value & PCC & p-value \\
    \hline
    Recall@100 & 0.0814 & 0.4425 & 0.1974 & 0.0606 \\
    NDCG & 0.3719 & 0.0002 &  0.3825 & 0.0002\\
    List AUC & 0.4353 & $1.6\times 10^{-5}$ & 0.1564 & 0.1386\\
    \hline
    WR@100 & 0.7879 & $1.9\times 10^{-20}$ & 0.8666 & $2.5\times 10^{-42}$ \\
    % WR$_{log}$@100 & 0.7359 & $9.5\times 10^{-17}$ & 0.8056 & $6.0\times 10^{-22}$ \\
    % WR$_{sqrt}$@100 & 0.7054 & $5.8\times 10^{-15}$ & 0.8281 & $4.2\times 10^{-24}$ \\
    % WR$_{square}$@100 & 0.6797 & $1.3\times 10^{-13}$ & 0.8515 & $1.1\times 10^{-26}$ \\
    \hline
    \hline
    \end{tabular}
    \label{tab:result-metric-correlation}
\end{table}

\subsection{Score Fusion Strategy}

% Ranking items solely based on a single estimated target cannot well match the online requirement, e.g., CTR does not account for conversion, while CTCVR can lead to too narrow coverage. 
% better balance the user preference diversity
% enhance the diversity of related products to be displayed
% which may require exposing more diverse items of interest to users (even if they may not buy)

To align with the online metric (OPU in our case) and better provide the service\footnote{It may require exposing more diverse items of interest to users including those they won't order. This might be why CTR is typically incorporated in score fusion formulas.}, ranking systems in e-commerce platforms typically fuse multiple scores to form the indicator for item ranking. A historically employed formula for score fusion combines\footnote{One could include a price term to balance market scale measured by order quantity and gross merchandise value. We mainly consider OPU hence omit it for clarity.} estimated CTR and CTCVR multiplicatively \cite{ma2018entire}: 
\begin{equation} \label{eq:score-formula1}
    Score = CTR^{\alpha} \times CVR^{\beta},
\end{equation}
which can be equivalently formulated as $CTR^{\alpha-\beta} \times CTCVR^{\beta}$. 
In our experiments, we investigated fusing CTR and CTCVR additively: 
\begin{equation} \label{eq:score-formula2}
    Score = \alpha \cdot CTR + \beta \cdot CTCVR,
\end{equation}
which we found to be more effective. Besides, we tried to further additively incorporate semantic relevance indicators from other teams, which also seems valuable. 

% Fusion formulas can be categorized based on their elements (score-based or index-based) and their functions (additive or multiplicative). Multiplicative formulas, like $CTR^{\alpha}\times CVR^{\beta}$, are common in causal frameworks \cite{ma2018entire,wang2022escm2,zhu2023dcmt}. Additive formulas, such as $\alpha \cdot CTR + \beta \cdot CTCVR$, offer an alternative approach. For index-based formulas, scores are replaced with indices derived from rankings based on each score. 

% Additionally, beyond interaction-driven scores, we incorporate experience-driven scores like semantic relevance into the fusion process. We explored two strategies to combine interaction-driven and experience-driven scores. The first strategy involves a weighted sum, blending these scores directly. The second strategy discretizes relevance scores into three levels (0, 1, and 2), representing low to high relevance. Then it applies a weighted sum of the interaction-driven scores and these discretized relevance levels. 

We search the best weight parameter (e.g., $\alpha$ and $\beta$) for all these different fusion formulations w.r.t. offline WR@K for ESMM and ResFlow, and for each of the best, we conduct a two-week online A/B test. The results are shown in Table \ref{tab:score-formula}, where we use ESMM with CTR$^1$$\times$CVR$^1$, i.e., directly using CTCVR as the ranking indicator, as the baseline, and report the relative performances. 
We see that: 1) ResFlow with CTCVR outperforms ESMM with CTCVR; 2) score fusions generally lead to better performance than directly using CTCVR; 3) ResFlow with optimal multiplicative fusion outperforms that of ESMM; 4) both ESMM and ResFlow show better performance with additive formulas; 5) ResFlow with optimal additive fusion outperforms that of ESMM; 6) best-performing formulas\footnote{Note that CTR$^{\text{-0.2}}$$\times$CTCVR is equivalent to CTR$^{\text{0.8}}$$\times$CVR$^{\text{1.0}}$.} tend to enhance CTR relative to CVR or CTCVR; 7) adding an optimally weighted relevance score (RS) leads to a decrease in OPU though improves the bad case rate (the rate of top-ranked items mismatching user query, from another team); 8) adding an optimally weighted relevance level (RL, three-level discretized relevance score) degenerates OPU slightly, while notably improving the bad case rate. 

We consider that multiplicative fusion may be too sensitive to extreme values, e.g., if one score (e.g., CTCVR) is very small, the others could lose their voting rights, while additive fusion provides milder control.  Relevance level helps steadily enhance item relevance. 

% magnify discrepancies in individual scores

\subsection{Online Performance}

With the best score fusion formulas for each, we compared 3-target ResFlow against 3-target ESMM with a two-week online A/B test. ResFlow achieved a 1.29\% uplift in OPU, a 0.88\% increase in gross merchandise value, a 0.84\% rise in the number of buyers, a 0.25\% increase in online CTR, and a 1.37\% uplift in online CTCVR, compared to ESMM, without an increase in the bad case rate and system latency, with average and 99th percentile latency being 110 ms and 147 ms respectively for ResFlow v.s. 110 ms and 146 ms for ESMM.

\section{Conclusion} \label{sec:conclusion}

We have proposed ResFlow, a lightweight multi-task learning framework that enables information transfer via inter-task residual connections, and demonstrated its general effectiveness in various application scenarios with extensive offline and online experiments. We have addressed key implementation challenges of deploying multi-task learning in applied ranking systems, including online-offline metric alignment and multi-score fusion strategies. ResFlow is now fully deployed in the pre-rank module of Shopee Search. 

% three paragraph: 
% 1. our model novelty
% 2. practice novelty, business value, audience, generality 
% 3. limitation and future work

\begin{acks}
    This research is supported in part by the National Natural Science Foundation of China (U22B2020).
    % the National Research Foundation Singapore and DSO National Laboratories under the AI Singapore Programme (AISG2-RP-2020-019); the RIE 2020 Advanced Manufacturing and Engineering Programmatic Fund (A20G8b0102), Singapore. 
\end{acks}

\bibliographystyle{ACM-Reference-Format}
\bibliography{main}


\begin{thebibliography}{40}

%%% the Web address.

\ifx \showCODEN    \undefined \def \showCODEN     #1{\unskip}     \fi
\ifx \showDOI      \undefined \def \showDOI       #1{#1}\fi
\ifx \showISBNx    \undefined \def \showISBNx     #1{\unskip}     \fi
\ifx \showISBNxiii \undefined \def \showISBNxiii  #1{\unskip}     \fi
\ifx \showISSN     \undefined \def \showISSN      #1{\unskip}     \fi
\ifx \showLCCN     \undefined \def \showLCCN      #1{\unskip}     \fi
\ifx \shownote     \undefined \def \shownote      #1{#1}          \fi
\ifx \showarticletitle \undefined \def \showarticletitle #1{#1}   \fi
\ifx \showURL      \undefined \def \showURL       {\relax}        \fi
% invisible to TeX
\providecommand\bibfield[2]{#2}
\providecommand\bibinfo[2]{#2}
\providecommand\natexlab[1]{#1}
\providecommand\showeprint[2][]{arXiv:#2}

\bibitem[Batista et~al\mbox{.}(2004)]%
        {batista2004study}
\bibfield{author}{\bibinfo{person}{Gustavo~EAPA Batista}, \bibinfo{person}{Ronaldo~C Prati}, {and} \bibinfo{person}{Maria~Carolina Monard}.} \bibinfo{year}{2004}\natexlab{}.
\newblock \showarticletitle{A study of the behavior of several methods for balancing machine learning training data}.
\newblock \bibinfo{journal}{\emph{ACM SIGKDD explorations newsletter}} \bibinfo{volume}{6}, \bibinfo{number}{1} (\bibinfo{year}{2004}), \bibinfo{pages}{20--29}.
\newblock


\bibitem[Beel et~al\mbox{.}(2013)]%
        {beel2013comparative}
\bibfield{author}{\bibinfo{person}{Joeran Beel}, \bibinfo{person}{Marcel Genzmehr}, \bibinfo{person}{Stefan Langer}, \bibinfo{person}{Andreas N{\"u}rnberger}, {and} \bibinfo{person}{Bela Gipp}.} \bibinfo{year}{2013}\natexlab{}.
\newblock \showarticletitle{A comparative analysis of offline and online evaluations and discussion of research paper recommender system evaluation}. In \bibinfo{booktitle}{\emph{Proceedings of the international workshop on reproducibility and replication in recommender systems evaluation}}. \bibinfo{pages}{7--14}.
\newblock


\bibitem[Caruana(1997)]%
        {caruana1997multitask}
\bibfield{author}{\bibinfo{person}{Rich Caruana}.} \bibinfo{year}{1997}\natexlab{}.
\newblock \showarticletitle{Multitask learning}.
\newblock \bibinfo{journal}{\emph{Machine learning}}  \bibinfo{volume}{28} (\bibinfo{year}{1997}), \bibinfo{pages}{41--75}.
\newblock


\bibitem[Chawla et~al\mbox{.}(2004)]%
        {chawla2004special}
\bibfield{author}{\bibinfo{person}{Nitesh~V Chawla}, \bibinfo{person}{Nathalie Japkowicz}, {and} \bibinfo{person}{Aleksander Kotcz}.} \bibinfo{year}{2004}\natexlab{}.
\newblock \showarticletitle{Special issue on learning from imbalanced data sets}.
\newblock \bibinfo{journal}{\emph{ACM SIGKDD explorations newsletter}} \bibinfo{volume}{6}, \bibinfo{number}{1} (\bibinfo{year}{2004}), \bibinfo{pages}{1--6}.
\newblock


\bibitem[Chen et~al\mbox{.}(2023a)]%
        {chen2023bias}
\bibfield{author}{\bibinfo{person}{Jiawei Chen}, \bibinfo{person}{Hande Dong}, \bibinfo{person}{Xiang Wang}, \bibinfo{person}{Fuli Feng}, \bibinfo{person}{Meng Wang}, {and} \bibinfo{person}{Xiangnan He}.} \bibinfo{year}{2023}\natexlab{a}.
\newblock \showarticletitle{Bias and debias in recommender system: A survey and future directions}.
\newblock \bibinfo{journal}{\emph{ACM Transactions on Information Systems}} \bibinfo{volume}{41}, \bibinfo{number}{3} (\bibinfo{year}{2023}), \bibinfo{pages}{1--39}.
\newblock


\bibitem[Chen et~al\mbox{.}(2023b)]%
        {chen2023clustered}
\bibfield{author}{\bibinfo{person}{Yizhou Chen}, \bibinfo{person}{Guangda Huzhang}, \bibinfo{person}{Anxiang Zeng}, \bibinfo{person}{Qingtao Yu}, \bibinfo{person}{Hui Sun}, \bibinfo{person}{Heng-Yi Li}, \bibinfo{person}{Jingyi Li}, \bibinfo{person}{Yabo Ni}, \bibinfo{person}{Han Yu}, {and} \bibinfo{person}{Zhiming Zhou}.} \bibinfo{year}{2023}\natexlab{b}.
\newblock \showarticletitle{Clustered Embedding Learning for Recommender Systems}. In \bibinfo{booktitle}{\emph{Proceedings of the ACM Web Conference 2023}}. \bibinfo{pages}{1074--1084}.
\newblock


\bibitem[Du et~al\mbox{.}(2022)]%
        {du2022hierarchical}
\bibfield{author}{\bibinfo{person}{Jing Du}, \bibinfo{person}{Lina Yao}, \bibinfo{person}{Xianzhi Wang}, \bibinfo{person}{Bin Guo}, {and} \bibinfo{person}{Zhiwen Yu}.} \bibinfo{year}{2022}\natexlab{}.
\newblock \showarticletitle{Hierarchical Task-aware Multi-Head Attention Network}. In \bibinfo{booktitle}{\emph{Proceedings of the 45th International ACM SIGIR Conference on Research and Development in Information Retrieval}}. \bibinfo{pages}{1933--1937}.
\newblock


\bibitem[Gallagher et~al\mbox{.}(2019)]%
        {gallagher2019joint}
\bibfield{author}{\bibinfo{person}{Luke Gallagher}, \bibinfo{person}{Ruey-Cheng Chen}, \bibinfo{person}{Roi Blanco}, {and} \bibinfo{person}{J~Shane Culpepper}.} \bibinfo{year}{2019}\natexlab{}.
\newblock \showarticletitle{Joint optimization of cascade ranking models}. In \bibinfo{booktitle}{\emph{Proceedings of the twelfth ACM international conference on web search and data mining}}. \bibinfo{pages}{15--23}.
\newblock


\bibitem[Gao et~al\mbox{.}(2022)]%
        {gao2022kuairand}
\bibfield{author}{\bibinfo{person}{Chongming Gao}, \bibinfo{person}{Shijun Li}, \bibinfo{person}{Yuan Zhang}, \bibinfo{person}{Jiawei Chen}, \bibinfo{person}{Biao Li}, \bibinfo{person}{Wenqiang Lei}, \bibinfo{person}{Peng Jiang}, {and} \bibinfo{person}{Xiangnan He}.} \bibinfo{year}{2022}\natexlab{}.
\newblock \showarticletitle{KuaiRand: An Unbiased Sequential Recommendation Dataset with Randomly Exposed Videos}. In \bibinfo{booktitle}{\emph{Proceedings of the 31st ACM International Conference on Information and Knowledge Management}} (Atlanta, GA, USA) \emph{(\bibinfo{series}{CIKM '22})}. \bibinfo{pages}{3953–3957}.
\newblock
\urldef\tempurl%
\url{https://doi.org/10.1145/3511808.3557624}
\showDOI{\tempurl}


\bibitem[He et~al\mbox{.}(2016)]%
        {he2016deep}
\bibfield{author}{\bibinfo{person}{Kaiming He}, \bibinfo{person}{Xiangyu Zhang}, \bibinfo{person}{Shaoqing Ren}, {and} \bibinfo{person}{Jian Sun}.} \bibinfo{year}{2016}\natexlab{}.
\newblock \showarticletitle{Deep residual learning for image recognition}. In \bibinfo{booktitle}{\emph{Proceedings of the IEEE conference on computer vision and pattern recognition}}. \bibinfo{pages}{770--778}.
\newblock


\bibitem[Huang et~al\mbox{.}(2013)]%
        {huang2013learning}
\bibfield{author}{\bibinfo{person}{Po-Sen Huang}, \bibinfo{person}{Xiaodong He}, \bibinfo{person}{Jianfeng Gao}, \bibinfo{person}{Li Deng}, \bibinfo{person}{Alex Acero}, {and} \bibinfo{person}{Larry Heck}.} \bibinfo{year}{2013}\natexlab{}.
\newblock \showarticletitle{Learning deep structured semantic models for web search using clickthrough data}. In \bibinfo{booktitle}{\emph{Proceedings of the 22nd ACM international conference on Information \& Knowledge Management}}. \bibinfo{pages}{2333--2338}.
\newblock


\bibitem[Huzhang et~al\mbox{.}(2021)]%
        {huzhang2021aliexpress}
\bibfield{author}{\bibinfo{person}{Guangda Huzhang}, \bibinfo{person}{Zhenjia Pang}, \bibinfo{person}{Yongqing Gao}, \bibinfo{person}{Yawen Liu}, \bibinfo{person}{Weijie Shen}, \bibinfo{person}{Wen-Ji Zhou}, \bibinfo{person}{Qing Da}, \bibinfo{person}{Anxiang Zeng}, \bibinfo{person}{Han Yu}, \bibinfo{person}{Yang Yu}, {et~al\mbox{.}}} \bibinfo{year}{2021}\natexlab{}.
\newblock \showarticletitle{AliExpress Learning-To-Rank: Maximizing online model performance without going online}.
\newblock \bibinfo{journal}{\emph{IEEE Transactions on Knowledge and Data Engineering}} (\bibinfo{year}{2021}).
\newblock


\bibitem[Jin et~al\mbox{.}(2022)]%
        {jin2022multi}
\bibfield{author}{\bibinfo{person}{Jiarui Jin}, \bibinfo{person}{Xianyu Chen}, \bibinfo{person}{Weinan Zhang}, \bibinfo{person}{Yuanbo Chen}, \bibinfo{person}{Zaifan Jiang}, \bibinfo{person}{Zekun Zhu}, \bibinfo{person}{Zhewen Su}, {and} \bibinfo{person}{Yong Yu}.} \bibinfo{year}{2022}\natexlab{}.
\newblock \showarticletitle{Multi-Scale User Behavior Network for Entire Space Multi-Task Learning}. In \bibinfo{booktitle}{\emph{Proceedings of the 31st ACM International Conference on Information \& Knowledge Management}}. \bibinfo{pages}{874--883}.
\newblock


\bibitem[Lee et~al\mbox{.}(2020)]%
        {lee2020residual}
\bibfield{author}{\bibinfo{person}{Janghyeon Lee}, \bibinfo{person}{Donggyu Joo}, \bibinfo{person}{Hyeong~Gwon Hong}, {and} \bibinfo{person}{Junmo Kim}.} \bibinfo{year}{2020}\natexlab{}.
\newblock \showarticletitle{Residual continual learning}. In \bibinfo{booktitle}{\emph{Proceedings of the AAAI Conference on Artificial Intelligence}}, Vol.~\bibinfo{volume}{34}. \bibinfo{pages}{4553--4560}.
\newblock


\bibitem[Li et~al\mbox{.}(2020)]%
        {li2020improving}
\bibfield{author}{\bibinfo{person}{Pengcheng Li}, \bibinfo{person}{Runze Li}, \bibinfo{person}{Qing Da}, \bibinfo{person}{An-Xiang Zeng}, {and} \bibinfo{person}{Lijun Zhang}.} \bibinfo{year}{2020}\natexlab{}.
\newblock \showarticletitle{Improving multi-scenario learning to rank in e-commerce by exploiting task relationships in the label space}. In \bibinfo{booktitle}{\emph{Proceedings of the 29th ACM International Conference on Information \& Knowledge Management}}. \bibinfo{pages}{2605--2612}.
\newblock


\bibitem[Lu and Li(2013)]%
        {lu2013deep}
\bibfield{author}{\bibinfo{person}{Zhengdong Lu} {and} \bibinfo{person}{Hang Li}.} \bibinfo{year}{2013}\natexlab{}.
\newblock \showarticletitle{A deep architecture for matching short texts}.
\newblock \bibinfo{journal}{\emph{Advances in neural information processing systems}}  \bibinfo{volume}{26} (\bibinfo{year}{2013}).
\newblock


\bibitem[Ma et~al\mbox{.}(2018b)]%
        {ma2018modeling}
\bibfield{author}{\bibinfo{person}{Jiaqi Ma}, \bibinfo{person}{Zhe Zhao}, \bibinfo{person}{Xinyang Yi}, \bibinfo{person}{Jilin Chen}, \bibinfo{person}{Lichan Hong}, {and} \bibinfo{person}{Ed~H Chi}.} \bibinfo{year}{2018}\natexlab{b}.
\newblock \showarticletitle{Modeling task relationships in multi-task learning with multi-gate mixture-of-experts}. In \bibinfo{booktitle}{\emph{Proceedings of the 24th ACM SIGKDD international conference on knowledge discovery \& data mining}}. \bibinfo{pages}{1930--1939}.
\newblock


\bibitem[Ma et~al\mbox{.}(2018a)]%
        {ma2018entire}
\bibfield{author}{\bibinfo{person}{Xiao Ma}, \bibinfo{person}{Liqin Zhao}, \bibinfo{person}{Guan Huang}, \bibinfo{person}{Zhi Wang}, \bibinfo{person}{Zelin Hu}, \bibinfo{person}{Xiaoqiang Zhu}, {and} \bibinfo{person}{Kun Gai}.} \bibinfo{year}{2018}\natexlab{a}.
\newblock \showarticletitle{Entire space multi-task model: An effective approach for estimating post-click conversion rate}. In \bibinfo{booktitle}{\emph{The 41st International ACM SIGIR Conference on Research \& Development in Information Retrieval}}. \bibinfo{pages}{1137--1140}.
\newblock


\bibitem[Maloof(2003)]%
        {maloof2003learning}
\bibfield{author}{\bibinfo{person}{Marcus~A Maloof}.} \bibinfo{year}{2003}\natexlab{}.
\newblock \showarticletitle{Learning when data sets are imbalanced and when costs are unequal and unknown}. In \bibinfo{booktitle}{\emph{ICML-2003 workshop on learning from imbalanced data sets II}}, Vol.~\bibinfo{volume}{2}. \bibinfo{pages}{2--1}.
\newblock


\bibitem[Misra et~al\mbox{.}(2016)]%
        {misra2016cross}
\bibfield{author}{\bibinfo{person}{Ishan Misra}, \bibinfo{person}{Abhinav Shrivastava}, \bibinfo{person}{Abhinav Gupta}, {and} \bibinfo{person}{Martial Hebert}.} \bibinfo{year}{2016}\natexlab{}.
\newblock \showarticletitle{Cross-stitch networks for multi-task learning}. In \bibinfo{booktitle}{\emph{Proceedings of the IEEE conference on computer vision and pattern recognition}}. \bibinfo{pages}{3994--4003}.
\newblock


\bibitem[Qin et~al\mbox{.}(2022)]%
        {qin2022rankflow}
\bibfield{author}{\bibinfo{person}{Jiarui Qin}, \bibinfo{person}{Jiachen Zhu}, \bibinfo{person}{Bo Chen}, \bibinfo{person}{Zhirong Liu}, \bibinfo{person}{Weiwen Liu}, \bibinfo{person}{Ruiming Tang}, \bibinfo{person}{Rui Zhang}, \bibinfo{person}{Yong Yu}, {and} \bibinfo{person}{Weinan Zhang}.} \bibinfo{year}{2022}\natexlab{}.
\newblock \showarticletitle{RankFlow: Joint Optimization of Multi-Stage Cascade Ranking Systems as Flows}. In \bibinfo{booktitle}{\emph{Proceedings of the 45th International ACM SIGIR Conference on Research and Development in Information Retrieval}}. \bibinfo{pages}{814--824}.
\newblock


\bibitem[Qin et~al\mbox{.}(2020)]%
        {qin2020multitask}
\bibfield{author}{\bibinfo{person}{Zhen Qin}, \bibinfo{person}{Yicheng Cheng}, \bibinfo{person}{Zhe Zhao}, \bibinfo{person}{Zhe Chen}, \bibinfo{person}{Donald Metzler}, {and} \bibinfo{person}{Jingzheng Qin}.} \bibinfo{year}{2020}\natexlab{}.
\newblock \showarticletitle{Multitask mixture of sequential experts for user activity streams}. In \bibinfo{booktitle}{\emph{Proceedings of the 26th ACM SIGKDD International Conference on Knowledge Discovery \& Data Mining}}. \bibinfo{pages}{3083--3091}.
\newblock


\bibitem[Rebuffi et~al\mbox{.}(2017)]%
        {rebuffi2017learning}
\bibfield{author}{\bibinfo{person}{Sylvestre-Alvise Rebuffi}, \bibinfo{person}{Hakan Bilen}, {and} \bibinfo{person}{Andrea Vedaldi}.} \bibinfo{year}{2017}\natexlab{}.
\newblock \showarticletitle{Learning multiple visual domains with residual adapters}.
\newblock \bibinfo{journal}{\emph{Advances in neural information processing systems}}  \bibinfo{volume}{30} (\bibinfo{year}{2017}).
\newblock


\bibitem[Schnabel et~al\mbox{.}(2016)]%
        {schnabel2016recommendations}
\bibfield{author}{\bibinfo{person}{Tobias Schnabel}, \bibinfo{person}{Adith Swaminathan}, \bibinfo{person}{Ashudeep Singh}, \bibinfo{person}{Navin Chandak}, {and} \bibinfo{person}{Thorsten Joachims}.} \bibinfo{year}{2016}\natexlab{}.
\newblock \showarticletitle{Recommendations as treatments: Debiasing learning and evaluation}. In \bibinfo{booktitle}{\emph{international conference on machine learning}}. PMLR, \bibinfo{pages}{1670--1679}.
\newblock


\bibitem[Tang et~al\mbox{.}(2020)]%
        {tang2020progressive}
\bibfield{author}{\bibinfo{person}{Hongyan Tang}, \bibinfo{person}{Junning Liu}, \bibinfo{person}{Ming Zhao}, {and} \bibinfo{person}{Xudong Gong}.} \bibinfo{year}{2020}\natexlab{}.
\newblock \showarticletitle{Progressive layered extraction (ple): A novel multi-task learning (mtl) model for personalized recommendations}. In \bibinfo{booktitle}{\emph{Proceedings of the 14th ACM Conference on Recommender Systems}}. \bibinfo{pages}{269--278}.
\newblock


\bibitem[Tao et~al\mbox{.}(2023)]%
        {tao2023task}
\bibfield{author}{\bibinfo{person}{Xuewen Tao}, \bibinfo{person}{Mingming Ha}, \bibinfo{person}{Qiongxu Ma}, \bibinfo{person}{Hongwei Cheng}, \bibinfo{person}{Wenfang Lin}, \bibinfo{person}{Xiaobo Guo}, \bibinfo{person}{Linxun Cheng}, {and} \bibinfo{person}{Bing Han}.} \bibinfo{year}{2023}\natexlab{}.
\newblock \showarticletitle{Task Aware Feature Extraction Framework for Sequential Dependence Multi-Task Learning}. In \bibinfo{booktitle}{\emph{Proceedings of the 17th ACM Conference on Recommender Systems}}. \bibinfo{pages}{151--160}.
\newblock


\bibitem[Wang et~al\mbox{.}(2022)]%
        {wang2022escm2}
\bibfield{author}{\bibinfo{person}{Hao Wang}, \bibinfo{person}{Tai-Wei Chang}, \bibinfo{person}{Tianqiao Liu}, \bibinfo{person}{Jianmin Huang}, \bibinfo{person}{Zhichao Chen}, \bibinfo{person}{Chao Yu}, \bibinfo{person}{Ruopeng Li}, {and} \bibinfo{person}{Wei Chu}.} \bibinfo{year}{2022}\natexlab{}.
\newblock \showarticletitle{Escm2: Entire space counterfactual multi-task model for post-click conversion rate estimation}. In \bibinfo{booktitle}{\emph{Proceedings of the 45th International ACM SIGIR Conference on Research and Development in Information Retrieval}}. \bibinfo{pages}{363--372}.
\newblock


\bibitem[Wang et~al\mbox{.}(2020)]%
        {wang2020cold}
\bibfield{author}{\bibinfo{person}{Zhe Wang}, \bibinfo{person}{Liqin Zhao}, \bibinfo{person}{Biye Jiang}, \bibinfo{person}{Guorui Zhou}, \bibinfo{person}{Xiaoqiang Zhu}, {and} \bibinfo{person}{Kun Gai}.} \bibinfo{year}{2020}\natexlab{}.
\newblock \showarticletitle{COLD: Towards the Next Generation of Pre-Ranking System}.
\newblock \bibinfo{journal}{\emph{Workshop on Deep Learning Practice for High-Dimensional Sparse Data in KDD 2020}} (\bibinfo{year}{2020}).
\newblock


\bibitem[Wu and Chang(2004)]%
        {wu2004aligning}
\bibfield{author}{\bibinfo{person}{Gang Wu} {and} \bibinfo{person}{Edward~Y Chang}.} \bibinfo{year}{2004}\natexlab{}.
\newblock \showarticletitle{Aligning boundary in kernel space for learning imbalanced dataset}. In \bibinfo{booktitle}{\emph{Fourth IEEE International Conference on Data Mining (ICDM'04)}}. IEEE, \bibinfo{pages}{265--272}.
\newblock


\bibitem[Wu et~al\mbox{.}(2022)]%
        {wu2022multi}
\bibfield{author}{\bibinfo{person}{Yiqing Wu}, \bibinfo{person}{Ruobing Xie}, \bibinfo{person}{Yongchun Zhu}, \bibinfo{person}{Xiang Ao}, \bibinfo{person}{Xin Chen}, \bibinfo{person}{Xu Zhang}, \bibinfo{person}{Fuzhen Zhuang}, \bibinfo{person}{Leyu Lin}, {and} \bibinfo{person}{Qing He}.} \bibinfo{year}{2022}\natexlab{}.
\newblock \showarticletitle{Multi-view multi-behavior contrastive learning in recommendation}. In \bibinfo{booktitle}{\emph{International Conference on Database Systems for Advanced Applications}}. Springer, \bibinfo{pages}{166--182}.
\newblock


\bibitem[Xi et~al\mbox{.}(2021)]%
        {xi2021modeling}
\bibfield{author}{\bibinfo{person}{Dongbo Xi}, \bibinfo{person}{Zhen Chen}, \bibinfo{person}{Peng Yan}, \bibinfo{person}{Yinger Zhang}, \bibinfo{person}{Yongchun Zhu}, \bibinfo{person}{Fuzhen Zhuang}, {and} \bibinfo{person}{Yu Chen}.} \bibinfo{year}{2021}\natexlab{}.
\newblock \showarticletitle{Modeling the sequential dependence among audience multi-step conversions with multi-task learning in targeted display advertising}. In \bibinfo{booktitle}{\emph{Proceedings of the 27th ACM SIGKDD Conference on Knowledge Discovery \& Data Mining}}. \bibinfo{pages}{3745--3755}.
\newblock


\bibitem[Yang and Hospedales(2017)]%
        {yang2017trace}
\bibfield{author}{\bibinfo{person}{Yongxin Yang} {and} \bibinfo{person}{Timothy Hospedales}.} \bibinfo{year}{2017}\natexlab{}.
\newblock \showarticletitle{Trace Norm Regularised Deep Multi-Task Learning}. In \bibinfo{booktitle}{\emph{International Conference on Learning Representations}}.
\newblock


\bibitem[Yu et~al\mbox{.}(2023)]%
        {yu2023low}
\bibfield{author}{\bibinfo{person}{Yu Yu}, \bibinfo{person}{Chao-Han~Huck Yang}, \bibinfo{person}{Jari Kolehmainen}, \bibinfo{person}{Prashanth~G Shivakumar}, \bibinfo{person}{Yile Gu}, \bibinfo{person}{Sungho Ryu}, \bibinfo{person}{Roger Ren}, \bibinfo{person}{Qi Luo}, \bibinfo{person}{Aditya Gourav}, \bibinfo{person}{I-Fan Chen}, {et~al\mbox{.}}} \bibinfo{year}{2023}\natexlab{}.
\newblock \showarticletitle{Low-rank adaptation of large language model rescoring for parameter-efficient speech recognition}.
\newblock \bibinfo{journal}{\emph{arXiv preprint arXiv:2309.15223}} (\bibinfo{year}{2023}).
\newblock


\bibitem[Zhang et~al\mbox{.}(2023b)]%
        {zhang2023adding}
\bibfield{author}{\bibinfo{person}{Lvmin Zhang}, \bibinfo{person}{Anyi Rao}, {and} \bibinfo{person}{Maneesh Agrawala}.} \bibinfo{year}{2023}\natexlab{b}.
\newblock \showarticletitle{Adding conditional control to text-to-image diffusion models}. In \bibinfo{booktitle}{\emph{Proceedings of the IEEE/CVF International Conference on Computer Vision}}. \bibinfo{pages}{3836--3847}.
\newblock


\bibitem[Zhang and Yang(2017)]%
        {zhang2017learning}
\bibfield{author}{\bibinfo{person}{Yu Zhang} {and} \bibinfo{person}{Qiang Yang}.} \bibinfo{year}{2017}\natexlab{}.
\newblock \showarticletitle{Learning sparse task relations in multi-task learning}. In \bibinfo{booktitle}{\emph{Proceedings of the AAAI Conference on Artificial Intelligence}}, Vol.~\bibinfo{volume}{31}.
\newblock


\bibitem[Zhang and Yang(2021)]%
        {zhang2021survey}
\bibfield{author}{\bibinfo{person}{Yu Zhang} {and} \bibinfo{person}{Qiang Yang}.} \bibinfo{year}{2021}\natexlab{}.
\newblock \showarticletitle{A survey on multi-task learning}.
\newblock \bibinfo{journal}{\emph{IEEE Transactions on Knowledge and Data Engineering}} \bibinfo{volume}{34}, \bibinfo{number}{12} (\bibinfo{year}{2021}), \bibinfo{pages}{5586--5609}.
\newblock


\bibitem[Zhang et~al\mbox{.}(2010)]%
        {zhang2010probabilistic}
\bibfield{author}{\bibinfo{person}{Yu Zhang}, \bibinfo{person}{Dit-Yan Yeung}, {and} \bibinfo{person}{Qian Xu}.} \bibinfo{year}{2010}\natexlab{}.
\newblock \showarticletitle{Probabilistic multi-task feature selection}.
\newblock \bibinfo{journal}{\emph{Advances in neural information processing systems}}  \bibinfo{volume}{23} (\bibinfo{year}{2010}).
\newblock


\bibitem[Zhang et~al\mbox{.}(2023a)]%
        {zhang2023rethinking}
\bibfield{author}{\bibinfo{person}{Zhixuan Zhang}, \bibinfo{person}{Yuheng Huang}, \bibinfo{person}{Dan Ou}, \bibinfo{person}{Sen Li}, \bibinfo{person}{Longbin Li}, \bibinfo{person}{Qingwen Liu}, {and} \bibinfo{person}{Xiaoyi Zeng}.} \bibinfo{year}{2023}\natexlab{a}.
\newblock \showarticletitle{Rethinking the Role of Pre-ranking in Large-scale E-Commerce Searching System}.
\newblock \bibinfo{journal}{\emph{arXiv preprint arXiv:2305.13647}} (\bibinfo{year}{2023}).
\newblock


\bibitem[Zheng and Wang(2022)]%
        {zheng2022survey}
\bibfield{author}{\bibinfo{person}{Yong Zheng} {and} \bibinfo{person}{David~Xuejun Wang}.} \bibinfo{year}{2022}\natexlab{}.
\newblock \showarticletitle{A survey of recommender systems with multi-objective optimization}.
\newblock \bibinfo{journal}{\emph{Neurocomputing}}  \bibinfo{volume}{474} (\bibinfo{year}{2022}), \bibinfo{pages}{141--153}.
\newblock


\bibitem[Zhu et~al\mbox{.}(2023)]%
        {zhu2023dcmt}
\bibfield{author}{\bibinfo{person}{Feng Zhu}, \bibinfo{person}{Mingjie Zhong}, \bibinfo{person}{Xinxing Yang}, \bibinfo{person}{Longfei Li}, \bibinfo{person}{Lu Yu}, \bibinfo{person}{Tiehua Zhang}, \bibinfo{person}{Jun Zhou}, \bibinfo{person}{Chaochao Chen}, \bibinfo{person}{Fei Wu}, \bibinfo{person}{Guanfeng Liu}, {et~al\mbox{.}}} \bibinfo{year}{2023}\natexlab{}.
\newblock \showarticletitle{DCMT: A Direct Entire-Space Causal Multi-Task Framework for Post-Click Conversion Estimation}. In \bibinfo{booktitle}{\emph{Proceedings of the39th International Conference on Data Engineering}}. \bibinfo{pages}{3113--3125}.
\newblock


\end{thebibliography}

\clearpage
\appendix

\section{More Experiment Results} \label{sec:app-a}

\begin{figure*}
  \centering
  \includegraphics[width=0.98\linewidth]{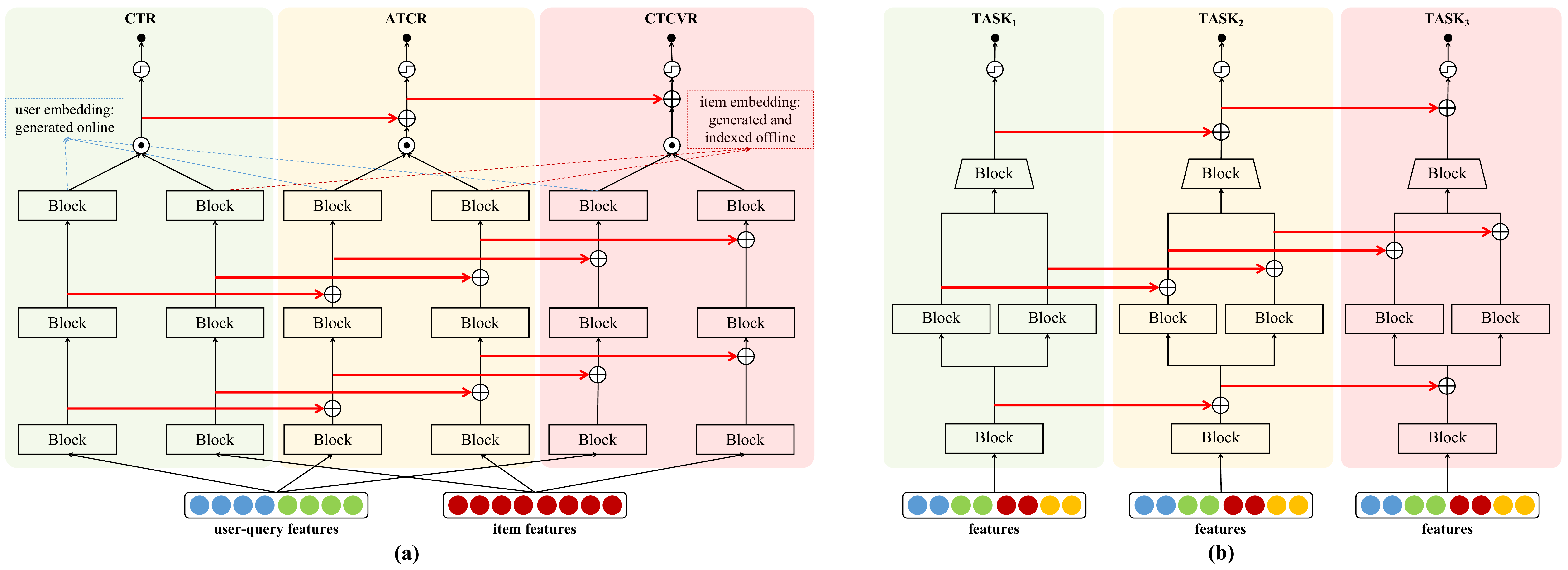}
  \vspace{-5pt}
  \caption{ResFlow with more complex architectures. (a) The illustration of ResFlow builds upon the twin-tower architecture. It is adopted as the backbone for all experiments on the Shopee dataset and our online deployment. (b) Another conceptual ResFlow architecture, where task networks have inner branches. Residual connections can be generally applied behind function blocks that have parameters, while we should be cautious about operations that have no or very few parameters (see Section~\ref{sec:twin-tower}).} % which turn them into residual learners and make their learning easier, 
  \label{fig:pre-rank-two-tower}
\end{figure*}

\begin{figure*}
  \centering
  \includegraphics[width=0.98\linewidth]{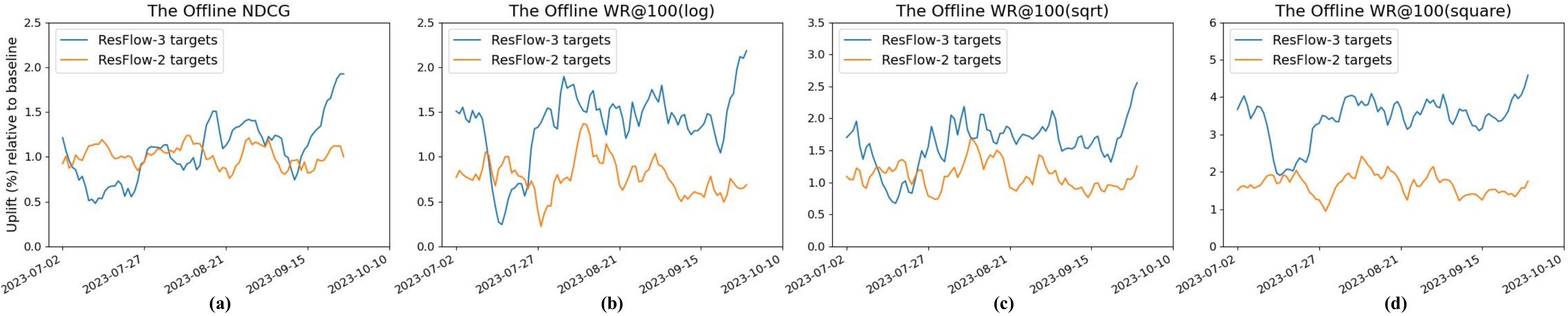}
  \caption{The uplift of more offline metrics of ResFlow relative to the formerly deployed method (ESMM) during the online A/B test. The curves are smoothed to better show the trend.  }
  \label{fig:online-experiment2}
\end{figure*}

\subsection{Residual Links in Twin-Tower Architecture} \label{sec:twin-tower}

For the twin-tower architecture, illustrated in Figure \ref{fig:pre-rank-two-tower}(a), which is typically employed in the pre-rank module, establishing residual connections is slightly different from that of the single-tower architecture. The main concern is where to put the residual links around the highest-level abstractions, i.e., before or after the inner product. We have tested different choices. The results are shown in Table~\ref{tab:two-tower-final-residual}. We can see that positioning the residual link before or after the inner product presents similar performances, while putting it after the inner product, i.e., building a residual connection on the logits, leads to slightly better performance. We consider that putting a residual link after the inner product would treat the inner product operation along with the last function blocks of the two towers as the final residual learner, which offers more degrees of freedom in learning and can leverage a higher degree of abstraction from the former task, whereas adding residual links before the inner product is also fine, which defines the last function block of each tower as a residual learner respectively. 
In contrast, adding residual links to both before and after the inner product degenerates the performance significantly. This should be because the residual link before the inner product will turn the last function blocks into residual learners respectively for the two towers, but meanwhile, the residual link after the inner product also makes the inner product a residual learner. However, there are no learnable parameters in the inner product operation, hence it lacks the expressiveness needed to be a residual learner, which would not ease the learning process but bring about more burdens to the lower levels. 
Positioning the residual link after the dot-product operation is adopted for ResFlow in Section \ref{sec:exp-e-commerce-perf} and in online deployment.

\subsection{Enforcing Non-Increase of Probability}

In sequentially dependent multi-task settings, e.g., "click" $\shortrightarrow$ "add-to-cart" $\shortrightarrow$ "order", when modeling the probabilities of progressing from the outset to each of the later stages, it is expected that the predicted probabilities to be non-increasing along the dependency chain since only when the former action takes place the latter could possibly happen. According to our experiments, such non-increase can naturally be learned by the task networks without additional regularizers in most cases, especially in e-commerce settings where the ground truth probability for each task shows significant disparities. Nevertheless, we investigated several regularization techniques for enforcing such a constraint, including: 
\begin{enumerate}[leftmargin=*]
    \item Penalizing the increase in probabilities via a regularization defined as, akin to the one used in AITM~\cite{xi2021modeling}:
        \begin{equation}
            L_{prob}=\sum_i\sum_{k=2}^K \max(\hat{y}_k^i - \hat{y}_{k-1}^i,0).
        \end{equation}    
    \item Penalizing residual logits exceeding zero:
        \begin{equation}
            L_{logit}=\sum_i\sum_{k=2}^K \max(logit_k^i,0). 
        \end{equation}    
    \item Mandating residual logits to be non-positive, by involving a $\min(logit,0)$ operator after the logit and before the sigmoid.      
    \item Generating the output probability via the multiplication of the output probability of the former task and the probability generated by a new-build task network, i.e., using the ESMM framework, to ensure the non-increase of probability. 
\end{enumerate}
The results are shown in Table~\ref{tab:calibration-loss}. We can see that mandating non-positive residual logits leads to the best results among the four. It leads to almost the same performance as the non-regularized version in most cases, while in some cases brings slight improvement. Other regularizers, however, may affect the performance. 

\begin{table}
    \centering
    \caption{Results of regularizers for enforcing non-increase of probability. W.O.: ResFlow without non-increase regularizer. M1: ResFlow + penalizing increases in probabilities. M2: ResFlow + penalizing positive residual logits. M3: ResFlow + mandating non-positive residual logits. M4: ESMM + feature residual. We tune the best reg weight for each of them.}
    \resizebox{.48\textwidth}{!}{
    \begin{tabular}{cccc}
    \hline
    \hline
    Method & Dataset (Task) & Reg Weight & MSE/AUC \\
    \hline
    % AITM-like & kuaishou & 1:1 & 1657.84 \\
    % AITM-like & kuaishou & 1:2 & 1662.45 \\
    % AITM-like & kuaishou & 1:0.5 & 1658.95 \\    
    % -max(logit,0) & kuaishou & 1:1 & 2234.38 \\
    % -max(logit,0) & kuaishou & 1:2 & 20258.95\\ 
    % -max(logit,0) & kuaishou & 1:0.5 &1660.26 \\ 
    W.O. & KuaiRand-Pure-S1 (Regression) & NA & 1658.44 \\
    M1 & KuaiRand-Pure-S1 (Regression) & 0.1 & 1658.45 \\
    M2 & KuaiRand-Pure-S1 (Regression) & 0.1 & 1656.72 \\
    M3 & KuaiRand-Pure-S1 (Regression) & NA & 1655.83 \\
    M4 & KuaiRand-Pure-S1 (Regression) & NA & 1673.55 \\
    % \hline
    % \hline
    % Method & Dataset & Reg Weight & AUC \\
    \hline
    W.O. & Shopee-3 & NA & 0.9101 \\
    M1 & Shopee-3 & 0.1 & 0.9093 \\
    M2 & Shopee-3 & 0.1 & 0.9087 \\
    M3  & Shopee-3 & NA & 0.9102 \\
    M4 & Shopee-3 & NA & 0.9085 \\
    \hline
    % AITM-like & AE-RU & 1:1 & 0.9128 \\
    % AITM-like & AE-RU & 1:2 & 0.9124 \\
    W.O. & AE-RU & NA & 0.9134 \\
    M1 & AE-RU & 0.5 & 0.9133 \\
    % AITM-like & AE-RU & 1:0.1 & 0.9131 \\
    % -max(logit,0) & AE-RU & 1:1 & 0.8817 \\
    % -max(logit,0) & AE-RU & 1:2 & 0.7176\\ 
    % -max(logit,0) & AE-RU & 1:0.5 & 0.8898 \\ 
    M2 & AE-RU  & 0.1 & 0.9097 \\
    M3  & AE-RU & NA & 0.9135 \\
    M4  & AE-RU & NA & 0.9083 \\
    \hline
    \hline
    \end{tabular}
    }
    \label{tab:calibration-loss}
\end{table}

\subsection{WR@K Variants} 

We have tried different variants of WR@K by replacing the weight term $W_k$ with, for example, $\log(W_k)$, $\sqrt{W_k}$, and $W_k^2$. The results are shown in Table \ref{tab:result-metric-correlation2} and Figure~\ref{fig:online-experiment2}. Comparing with Table \ref{tab:result-metric-correlation}, we can see that the original version of WR@K aligns with online OPU best. 

% This is probably because the manipulation of the weight terms leads to over-emphasizing either low-popularity items or high-popularity items. 

\subsection{More Ablation and Baseline Results}

We provide the ablation results of ResFlow on AliCCP in Table~\ref{tab:more-ablation}. We provide CTCVR AUC the results of more baselines, including the single-task model, MOE, and MMOE, in Table~\ref{tab:more-baselines1}. 

% More results and experiment details can be found in Appendix C and D.

%, \ref{tab:more-baselines2}, and \ref{tab:more-ablation}

% This should be because both choices provide a well-defined way of how the last function blocks of the two towers should work upon the residual links. 

\begin{table}
    \centering
    \caption{Pearson correlation coefficient (PCC) between offline metric uplift and online metric uplift.}
    \begin{tabular}{c|cc|cc}
    \hline
    \hline
    \multirow{2}{*}{} & \multicolumn{2}{c|}{2 Targets Uplift} & \multicolumn{2}{c}{3 Targets Uplift} \\
    & PCC & p-value & PCC & p-value \\
    \hline
    WR@100 & 0.7879 & $1.9\times 10^{-20}$ & 0.8666 & $2.5\times 10^{-42}$ \\
    \hline
    % NDCG & 0.3719 & 0.0002 &  0.3825 & 0.0002\\
    WR$_{log}$@100 & 0.7359 & $9.5\times 10^{-17}$ & 0.8056 & $6.0\times 10^{-22}$ \\
    WR$_{sqrt}$@100 & 0.7054 & $5.8\times 10^{-15}$ & 0.8281 & $4.2\times 10^{-24}$ \\
    WR$_{square}$@100 & 0.6797 & $1.3\times 10^{-13}$ & 0.8515 & $1.1\times 10^{-26}$ \\
    \hline
    \hline
    \end{tabular}
    \label{tab:result-metric-correlation2}
\end{table}

\begin{table}
    \caption{Ablation results of the highest-level residual connections in the twin-tower architecture on Shopee-2.}
    \begin{tabular}{ccccc}
    \hline
    \hline
     Strategy & CTR AUC & CTCVR AUC \\ 
    \hline     
     Before \& After &  $0.8573\pm0.0018$ & $0.8801\pm0.0037$ \\
     Before Dot-Product & $0.8661\pm0.0021$ & $0.9019\pm0.0022$ \\ 
     After Dot-Product & $\mathbf{0.8667\pm0.0014}$ & $\mathbf{0.9024\pm0.0010}$ \\
    \hline
    \hline
    \end{tabular}
    \label{tab:two-tower-final-residual}
\end{table}

\begin{table}
\caption{Ablation results of ResFlow on AliCCP.}
\begin{tabular}{ccccc}
\hline
\hline
Model                       &        CTCVR AUC     \\
\hline
NSE                         & $0.6238 \pm 0.0018$ \\
NSE + Feature Residual (FR) & $0.6482 \pm 0.0010$  \\
NSE + FR(H1-only)           & $0.6296 \pm 0.0021$ \\
NSE + FR(H2-only)           & $0.6417 \pm 0.0015$  \\
NSE + Logit Residual (LR)   & $0.6511 \pm 0.0013$ \\
ResFlow(NSE + FR + LR)      & $\mathbf{0.6642 \pm 0.0016}$ \\
\hline
ESMM                        & $0.6412 \pm 0.0031$  \\
ESMM + FR                   & $0.6503 \pm 0.0028$  \\
ESMM + FR + LR              & $0.6487 \pm 0.0043$  \\
\hline
\hline
\end{tabular}
\label{tab:more-ablation}
\end{table}

\begin{table}
    \caption{More AUC results of the CTCVR estimation task on offline e-commercial datasets.}
    \begin{tabular}{ccccc}
    \hline
    \hline
    Datasets & Single Task & MOE & MMOE \\
    \hline
     S0 &0.609 $\pm$ 0.005& 0.624 $\pm$ 0.002 & 0.628 $\pm$ 0.001  \\
     S1 &0.609 $\pm$ 0.003& 0.620 $\pm$ 0.002 & 0.622 $\pm$ 0.001 \\
     S0\&S1 & 0.610 $\pm$ 0.002 & 0.629 $\pm$ 0.002 & 0.636 $\pm$ 0.002 \\
    AliCCP &0.613 $\pm$ 0.003& 0.637 $\pm$ 0.001 & 0.640 $\pm$ 0.001 \\
    \hline
    AE-ES  &0.844 $\pm$ 0.003& 0.871 $\pm$ 0.001 & 0.872 $\pm$ 0.003 \\
    AE-FR  &0.834 $\pm$ 0.003& 0.850 $\pm$ 0.003 & 0.851 $\pm$ 0.001 \\
    AE-NL  &0.813 $\pm$ 0.002& 0.831 $\pm$ 0.001 & 0.847 $\pm$ 0.003 \\
    AE-US  &0.812 $\pm$ 0.004& 0.844 $\pm$ 0.002 & 0.851 $\pm$ 0.002 \\
    AE-RU  & 0.862 $\pm$ 0.004& 0.876 $\pm$ 0.003 & 0.886 $\pm$ 0.003  \\
    \hline
    Shopee-2 & 0.859 $\pm$ 0.003& 0.851 $\pm$ 0.001 & 0.862 $\pm$ 0.002   \\
    Shopee-3 &0.859 $\pm$ 0.003& 0.876 $\pm$ 0.002 & 0.876 $\pm$ 0.003 \\
    \hline
    \hline
    \end{tabular}
    \label{tab:more-baselines1}
\end{table}

$ $ \newline
\vspace{10pt}
$ $ \newline
\vspace{10pt}
$ $ \newline

\end{document}